%% file: pengjie-tois-conversation.tex
\documentclass[acmsmall,natbib,screen]{acmart}

\input{packages}

\input{definitions}

\AtBeginDocument{%
  \providecommand\BibTeX{{%
    \normalfont B\kern-0.5em{\scshape i\kern-0.25em b}\kern-0.8em\TeX}}}

\setcopyright{acmcopyright}
\copyrightyear{2020}
\acmYear{2020}
\acmDOI{xx.xxxx/xxxxxxx.xxxxxxx}
\acmPrice{15.00}
\acmISBN{978-1-4503-9999-9/18/06}
\acmJournal{TOIS}

\begin{document}

\title{Conversations with Search Engines: SERP-based Conversational Response Generation}

\input{authors}

\renewcommand{\shortauthors}{Ren et al.}

\input{sections/00-abstract}



\maketitle

\input{sections/01-introduction}
\input{sections/02-dataset}
\input{sections/03-model}
\input{sections/04-experimentsetup}
\input{sections/05-results}
\input{sections/06-relatedwork}
\input{sections/07-conclusion}

\section*{Code and data}
The \ac{SaaC} dataset and the code of all the methods used for comparison in this paper are shared at \url{https://github.com/PengjieRen/CaSE-1.0}.

\begin{acks}
We thank our anonymous reviewers and the guest editors for helpful feedback.
This work was partially supported by 
the Netherlands Organisation for Scientific Research (NWO) under project nr CI-14-25,
the NWO Innovational Research Incentives Scheme Vidi (016.Vidi.189.039),
the NWO Smart Culture - Big Data / Digital Humanities (314-99-301),
the H2020-EU.3.4. - SOCIETAL CHALLENGES - Smart, Green And Integrated Transport (814961),
the Google Faculty Research Awards program,
the Natural Science Foundation of China (61972234, 61902219, 61672324, 61672322, 62072279), 
the Key Scientific and Technological Innovation Program of Shandong Province (2019JZZY010129), 
the Tencent AI Lab Rhino-Bird Focused Research Program (JR201932), 
the Fundamental Research Funds of Shandong University,
the National Key R\&D Program of China (2020YFB1406700),
and the Innovation Center for AI (ICAI).

All content represents the opinion of the authors, which is not necessarily shared or endorsed by their respective employers and/or sponsors.
\end{acks}

\bibliographystyle{ACM-Reference-Format}
\bibliography{bibtex}

\end{document}

%% file: packages.tex
\usepackage{booktabs} 
\usepackage[skip=0pt]{caption}
\usepackage{multirow}
\usepackage{bm}
\usepackage{balance}
\usepackage{algorithm}
\usepackage{algorithmic}
\usepackage{lineno}
\usepackage{subfig}
\usepackage[inline]{enumitem}
\usepackage{acronym}
\usepackage{pifont}
\usepackage{array}
\usepackage{graphicx}
\usepackage{tikz}
\usepackage{float}

%% file: definitions.tex
\AtBeginDocument{%
  \providecommand\BibTeX{{%
    \normalfont B\kern-0.5em{\scshape i\kern-0.25em b}\kern-0.8em\TeX}}}
\acrodef{BBC}{Background-Based Conversations}
\acrodef{DMFS}{Dual Multi-granularity Feature Selection}
\acrodef{PPG}{Prior-aware Pointer Generator}
\acrodef{CPU}{conversation \& passage understanding}
\acrodef{RPS}{relevant passage selection}
\acrodef{STI}{supporting token identification}
\acrodef{RG}{response generation}
\acrodef{CaSE}{\textbf{C}onvers\textbf{a}tions with \textbf{S}earch \textbf{E}ngines}
\acrodef{CCCE}{Confidence-Critical Cross Entropy}
\acrodef{CACE}{Copy-Adaptive Cross Entropy}
\acrodef{CS}{Conversational Search}
\acrodef{CA}{Conversational Agent}
\acrodef{SaaC}{Search as a Conversation}
\acrodef{CoQA}{Conversational Question Answering}
\acrodef{MultiWOZ}{Multi-Domain Wizard-of-Oz}
\acrodef{MRC}{machine reading comprehension}
\acrodef{QA}{question answering}
\acrodef{QuAC}{Question Answering in Context}
\acrodef{SERP}{search engine result page}
\acrodef{TDS}{task-oriented dialogue system}
\acrodef{KB}{knowledge base}
\acrodef{KBQA}{question answering over knowledge base}

\newcommand{\cmark}{\ding{51}}%
\newcommand{\xmark}{\ding{55}}%
\newcommand{\tabincell}[2]{\begin{tabular}{@{}#1@{}}#2\end{tabular}}  

\DeclareMathOperator{\attention}{attention}
\DeclareMathOperator{\softmax}{softmax}
\DeclareMathOperator{\mlp}{mlp}

\newcommand{\squeeze}{\vspace*{-2mm}}

\allowdisplaybreaks

%% file: authors.tex

\author{Pengjie Ren}
\affiliation{
 \institution{Shandong University \& University of Amsterdam}
 \city{Qingdao \& Amsterdam}
 \country{China \& The Netherlands}
}
\email{p.ren@uva.nl}

\author{Zhumin Chen$^*$}
\orcid{}
\affiliation{%
\institution{Shandong University}
\city{Qingdao}
\country{China}
}
\email{chenzhumin@sdu.edu.cn}

\author{Zhaochun Ren$^*$}
\orcid{}
\affiliation{%
\institution{Shandong University}
\city{Qingdao}
\country{China}
}
\email{zhaochun.ren@sdu.edu.cn}

\author{Evangelos Kanoulas}
\affiliation{
 \institution{University of Amsterdam}
 \city{Amsterdam}
 \country{The Netherlands}
}
\email{e.kanoulas@uva.nl}

\author{Christof Monz}
\affiliation{
 \institution{University of Amsterdam}
 \city{Amsterdam}
 \country{The Netherlands}
}
\email{c.monz@uva.nl}

\author{Maarten de Rijke}
\orcid{0000-0002-1086-0202}
\affiliation{
 \institution{University of Amsterdam \& Ahold Delhaize}
 \city{Amsterdam}
 \country{The Netherlands}
}
\email{m.derijke@uva.nl}

\thanks{$^*$corresponding authors}


%% file: sections/00-abstract.tex

\begin{abstract}
In this paper, we address the problem of answering complex information needs by conducting \emph{conversations with search engines}, in the sense that users can express their queries in natural language, and directly receive the information they need from a short system response in a conversational manner.
Recently, there have been some attempts towards a similar goal, e.g., studies on \acp{CA} and \ac{CS}.
However, they either do not address complex information needs in search scenarios, or they are limited to the development of conceptual frameworks and/or laboratory-based user studies.

We pursue two goals in this paper:
\begin{enumerate*}
\item the creation of a suitable dataset, the \ac{SaaC} dataset, for the development of pipelines for conversations with search engines, and
\item the development of a state-of-the-art pipeline for conversations with search engines, \ac{CaSE}, using this dataset.
\end{enumerate*}
\ac{SaaC} is built based on a multi-turn conversational search dataset, where we further employ workers from a crowdsourcing platform to summarize each relevant passage into a short, conversational response.
\ac{CaSE} enhances the state-of-the-art by introducing a supporting token identification module and a prior-aware pointer generator, which enables us to generate more accurate responses.

We carry out experiments to show that \ac{CaSE} is able to outperform strong baselines.
We also conduct extensive analyses on the \ac{SaaC} dataset to show where there is room for further improvement beyond \ac{CaSE}.
Finally, we release the \ac{SaaC} dataset and the code for \ac{CaSE} and all models used for comparison to facilitate future research on this topic.
\end{abstract}

%% file: sections/01-introduction.tex

\section{Introduction}

As we surround ourselves with a range of mobile devices, e.g., smartphones, smartwatches, which only have small screens or even no screen at all, search is increasingly performed in a conversational manner.\footnote{\url{https://www.thinkwithgoogle.com/consumer-insights/personal-needs-search-trends/}}
Despite this development, for complex information needs where a user's intent may be unclear or where it is not obvious what the single best direct answer should be (if any), complex \acp{SERP} are still the dominant format to present results to users.
\acp{SERP} are typically characterized by a diverse set of snippets, usually  grouped along vertical dimensions and/or by modality, which is far from our natural mode of communication through conversations.
\begin{figure}[h]
\centering
\includegraphics[width=0.8\columnwidth]{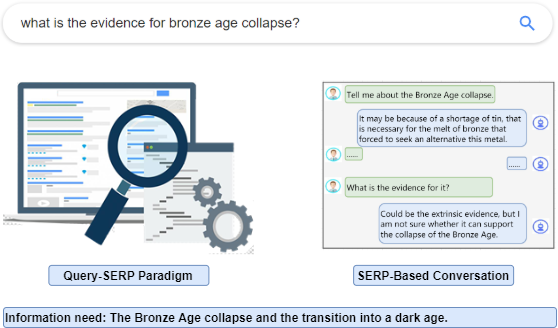}
\caption{Search using a traditional SERP vs.\ a conversation with a search engine, illustrated using the information need \emph{The Bronze Age collapse and the transition into a dark age}.
}
\label{01-1}
\end{figure}
%
Hence, even when we interact with search engines, the more natural mode of interaction, instead of a complex \ac{SERP} is conversational in nature~\citep{Croft:1987:IRN:35053.35054,Radlinski:2017:TFC:3020165.3020183,belkin1980anomalous,belkin1995cases}.
Figure~\ref{01-1} illustrates the difference given the information need ``\emph{The Bronze Age collapse and the transition into a dark age}.''
In a \emph{traditional \ac{SERP}} scenario, we would use keywords to express our information need.
For each query, we issue the keywords to a search engine and receive a \ac{SERP} with a ranked list of results, possibly with snippets, in return. 
Then, we go through the list and find the information we need from the relevant snippets and/or results.
If not, we reformulate our query and this process is repeated until our information need is satisfied.
Alternatively, we can fulfill our information need through a \emph{conversation with the search engine}.
To achieve this, there are several aspects to be considered, e.g., asking clarifying questions~\cite{Aliannejadi:2019:ACQ:3331184.3331265,Hamedwww2020}, conversational search and recommendation~\cite{cast2019,10.1145/3397271.3401130,10.1145/2939672.2939746,10.1145/3209978.3210002}, chitchats~\cite{wu-yan-2018-deep}, user intent understanding~\cite{10.1145/3295750.3298924}, feedback analysis~\cite{10.1145/3308558.3313661}, conversation management~\cite{TRIPPAS2020102162}, and conversational question answering~\cite{reddy-etal-2019-coqa}, etc.


In this work, we study conversational question answering in search scenarios, i.e., SERP-based conversational response generation, in the sense that we would express our need in natural language and we would directly receive the information we need in a short system response which is a summary of relevant results listed in the \ac{SERP} in a conversational manner.
Most studies on conversational question answering are based on knowledge bases (a.k.a \ac{KBQA})~\cite{yin-etal-2016-neural-generative} or a background passage (a.k.a \ac{MRC})~\citep{rajpurkar-etal-2016-squad,choi-etal-2018-quac}.
In these scenarios, users have well-specified and specific information needs and their queries are mostly factoid questions that can be answered by a relatively short text span (entity mentions, etc.) extracted from the given background knowledge (knowledge base, documents, etc.).
Several previous studies address complex user queries, e.g., on the QA over knowledge bases task.
However, the complex queries in these studies refer to compositional questions that involve joining multiple relations~\cite{DBLP:conf/aaai/SahaPKSC18}.
They are still factoid questions that can be answered by entity mentions or triples from knowledge bases.
In contrast, search engines cater for much broader needs/queries, which results in far more complex user queries and search scenarios.
Besides, despite the fact that some complex information needs can be answered by extracting a short span from \acp{SERP} in a single interaction, many require a deep understanding of multiple \acp{SERP} and multiple interactions are needed in order to achieve all the information needs, e.g., the one shown in Figure~\ref{01-1}.
There are some studies towards conversational question answering in search scenarios, but they all have critical limitations.
For instance, the CAsT dataset\footnote{\url{http://www.treccast.ai/}} only provides ground truth passages as answers and does not have conversational responses.
And the MS MARCO \ac{QA} dataset~\citep{DBLP:conf/nips/NguyenRSGTMD16} is single-turn and when there is no answer, it just leaves the system response blank, which is sufficient for training models but not suitable for evaluation. 
Hence, these datasets are not sufficient to support the development of conversations with search engines.

To address this gap we pursue two goals:
\begin{enumerate*}
\item the creation of a suitable dataset for the development of pipelines for SERP-based conversational response generation, and
\item the development of state-of-the-art baseline components that make up such a pipeline using this dataset.
\end{enumerate*}
The dataset, called \acfi{SaaC}, is developed in a Wizard-of-Oz fashion~\citep{bernsen2012designing}.
We simulate users based on conversational queries from the CAsT dataset.
Then, we employ online workers (a.k.a., ``wizards'') to play the role of the system.
The wizards have access to \acp{SERP} from which they can get useful information to respond to the user queries.
We ask the wizards to find supporting sentences from results/snippets on a \ac{SERP} and summarize them into short conversational responses. 
When there is no direct answer, we ask the wizards to generate something that is likely relevant, e.g., ``It could be the extrinsic evidence, but I am not sure \ldots'', or interesting to the user, e.g., ``I have no idea about how melatonin was discovered. But I can tell you that \ldots''.

As to our second main goal in this paper, using the \ac{SaaC} dataset for the development of pipelines to support conversations with search engines, 
we devise a modularized multi-task learning framework, called \acfi{CaSE}.
\ac{CaSE} decomposes conversations with search engines into four sub-tasks: 
\begin{enumerate*}
\item conversation \& passage understanding (\acs{CPU}\acused{CPU}), 
\item \ac{RPS}, 
\item \ac{STI}, and 
\item \ac{RG}.
\end{enumerate*}
\ac{CPU} is a module aiming at understanding and encoding conversations and passages.
\ac{RPS} then finds relevant passages based on the encoded representations from the \ac{CPU}.
\ac{STI} further identifies supporting tokens that are eventually used in the response.
Finally, \ac{RG} generates the response based on the output from the above three modules.

Because there are no ground truth labels for \ac{STI} to define a supervised learning loss, we present a weakly-supervised \ac{CCCE} learning loss based on the intuition that the overlapping tokens between the ground truth responses and the passages are more likely supporting tokens than the non-overlapping ones, especially overlapping larger rare tokens.
In order to make \ac{CaSE} generate more accurate responses, we propose a \ac{PPG} to implement \ac{RG} by considering the passage and token probabilities from \ac{RPS} and \ac{STI} as priors so that the generated responses are expected to be more accurate by including supporting tokens from relevant passages.

We conduct experiments to:
\begin{enumerate*}
\item compare the performance of state-of-the-art methods from related tasks to our \ac{CaSE}, 
\item understand the contribution of the four sub-tasks in \ac{CaSE}, and 
\item identify room for further improvement on \ac{SaaC} beyond \ac{CaSE}.
\end{enumerate*}

To sum up, the contributions of this work are as follows:
\begin{itemize}
\item We introduce the task of conversations with search engines and build a new dataset, the \ac{SaaC} dataset.
\item We decompose the task into four sub-tasks (\ac{CPU}, \ac{RPS}, \ac{STI}, \ac{RG}), and propose a modularized \ac{CaSE} model that uses a weakly-supervised \ac{CCCE} loss to identify supporting tokens and \ac{PPG} to encourage generating more accurate responses.
\item We conduct extensive experiments to show the effectiveness of \ac{CaSE} and identify room for further improvements on conversations with search engines.
\end{itemize}

\noindent%
In Section~\ref{s02} we describe the \ac{SaaC} dataset and its creation. 
In Section~\ref{03-model} we describe the \ac{CaSE} model.
Section~\ref{section:ExperimentalSetup} details our experimental setup and the outcomes of our experiments are presented in Section~\ref{section:ExperimentalResults}.
Related work is discussed in Section~\ref{section:RelatedWork} and we conclude in Section~\ref{section:conclusion}.

%% file: sections/02-dataset.tex

\section{The \NoCaseChange{SaaC} Dataset}
\label{s02}

Our goal is to create a dataset containing interactions between the users and a search engine where they communicate through conversations to achieve search tasks.
Instead of creating a new dataset from scratch, we build such a dataset based on the TREC CAsT dataset.
In this section we describe the stages involved in creating the \ac{SaaC} dataset.

\subsection{Collecting Conversational Queries}
TREC CAsT~\citep{cast2019} has already built a collection of conversational query sequences, so we reuse their data to reduce development cost.
Here, we briefly recall the process used in collecting the TREC CAsT data.

\begin{figure}[t]
\centering
\includegraphics[width=.7\columnwidth]{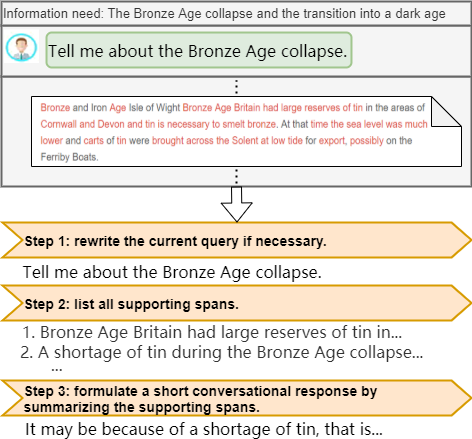}

\vspace*{2mm}
\caption{The process of building the \ac{SaaC} dataset.}
\label{02-1}
\end{figure}

The topics in the CAsT dataset are collected from a combination of previous TREC topics (Common Core~\cite{Kanoulas2009}, Session Track~\cite{Kanoulas2011}, etc.), MS MARCO Conversational Sessions, and the CAsT organizers~\citep{cast2019}.
The organizers ensured that the information needs are complex
(requiring multiple rounds of elaboration), diverse (across different information categories), open-domain (not requiring expert domain knowledge to access), and mostly answerable (with sufficient coverage in the passage collection).
A description of an example topic is shown in Figure~\ref{02-1} (top half).
Then, the TREC CAsT organizers created sequences of conversational queries for each turn. 
They started with the description of the topic and manually formulated the first conversational query.
After that, they formulated follow-up conversational queries by introducing coherent transitions, e.g., coreference and omission.
For example, ``Tell me about the Bronze Age collapse. \ldots\ What is the evidence for it? (introducing coreference).'', or ``What is a physician's assistant? \ldots\ What's the average starting salary in the UK? \ldots\ What about in the US? (introducing omission).''
There is a constraint that later conversational queries only depend on the previous queries, but not on system responses.
We will discuss this constraint later.
The reader is referred to \citep{cast2019} for a more detailed account of the creation of the TREC CAsT dataset.

\subsection{Collecting Candidate Passages}

For our dataset creation, we combine three standard TREC collections: MARCO Ranking passages, Wikipedia (TREC CAR), and News (Washington Post) as the passage collection.
To introduce more complex passages and meanwhile achieve higher recall for the current query, we follow \citet{ilpscast2019} to extend the current query by extracting words that capture relevant information from previous turns and add them to the query of the current turn.
Next, we use standard query likelihood with Dirichlet smoothing and RM3 relevance feedback as the ranking model to retrieve the top 10 candidate passages (if the ground truth passage is within the top 10, otherwise, we retrieve the top 9 and manually add the ground truth passage).
Note that although we rewrote the current query to make it self-contained (which will be detailed in the next subsection), we did not use the rewritten queries when preparing the candidate passages in order to stay close to practical search engines. 
Finally, we randomly shuffle the top 10 candidate passages to eliminate position bias.

\begin{table*}[t]
\centering
\caption{Comparison of conversational datasets on search, chitchat and question answering.}
\label{t02-1}
\setlength{\tabcolsep}{1pt}
\begin{tabular}{l c c c c}
\toprule
Dataset                          & \tabincell{c}{Multiple\\turns}              & \tabincell{c}{Natural language\\query} & \tabincell{c}{Multiple\\sources} & \tabincell{c}{Abstractive\\response}  \\
\midrule
Holl-e~\citep{moghe2018towards}, WoW~\citep{dinan2018wizard} & {\cmark}                     & {\cmark}             & {\xmark}          & {\cmark}                             \\
QuAC~\citep{choi-etal-2018-quac}, CoQA~\citep{reddy-etal-2019-coqa}                   & {\cmark}                     & {\xmark}              & {\xmark}          & {\xmark}                        \\
CAsT~\citep{cast2019} & \cmark & \cmark & \cmark & \xmark \\
MS MARCO~\citep{DBLP:conf/nips/NguyenRSGTMD16}                        & {\xmark}                      & {\textbf{?}}             & {\cmark}         & {\cmark}                                    \\
SaaC (this paper) & {\cmark}                     & {\cmark}             & {\cmark}         & {\cmark}                 \\\bottomrule
\end{tabular}
\squeeze\squeeze
\end{table*}

\subsection{Collecting Conversational Responses}
We employ online workers from Amazon Mechanical Turk (MTurk)\footnote{\url{https://www.mturk.com/}} to collect conversational responses in a Wizard-of-Oz fashion, where we ask the workers to play the role of the system and write responses based on the provided passages, as shown in Figure~\ref{02-1}.
Specifically, we first rewrite the queries if necessary and we require that the rewritten queries should be self-contained (step 1).
Then, the workers need to list all supporting spans from the passages that contain facts to help generate the responses (step 2).
The supporting spans are kept the same as they are in the passages.
Finally, we ask the workers to summarize the supporting spans into short, conversational responses.
Other requirements include:
\begin{enumerate*}
\item make sure the responses are case sensitive and grammatically correct; 
\item avoid using the spans directly in the responses without summarization; and
\item talk about secondary relevant information or information that could be interesting to the users, when no answer can be found in the passages.
\end{enumerate*}
We take three actions to guarantee that the collected data meets our requirements: 
First, we only employ high-quality master workers of MTurk who are required to continue to pass the statistical monitoring to retain the qualification. 
Second, we provide detailed guidelines, which also include several good and bad examples with detailed explanations.
Third, we manually checked each annotation by ourselves to guarantee the agreement between different workers, as it is hard to automatically check the quality of these responses written in natural language.
We make sure they all meet the given requirements:
\begin{enumerate*}
\item The queries are answered directly with further explanations;
\item The responses contain everything that is needed to answer the queries;
\item The supporting spans are summarized instead of being used directly in the responses;
\item The responses are grammatically and syntactically correct, and case sensitive.
\end{enumerate*}

In total, 28 master MTurk workers participated in our task. 
The number of annotations from each worker ranges from 1 to 244.
Table~\ref{t02-1} shows that while \ac{SaaC} shares some characteristics with existing datasets, it also has its own unique characteristics, tailored to support research on conversations with search engines.
For example, Holl-e and WoW contain multi-turn conversations, however they target chitchat.
QuAC and CoQA have multi-turn conversations, but the passages are all Wikipedia articles and fixed for all conversation turns; the responses are extractive text spans from Wikipedia articles.
The datasets listed above do not target search scenarios, where the passages are from various sources, e.g., blogs, Encyclopaedia, social media, and there are queries that cannot be answered properly with extractive text spans.
CAsT is for search scenarios, but it does not have abstractive responses.
Out of all the datasets available at the time of writing, MS MARCO is the closest to \ac{SaaC}.
However, MS MARCO is single-turn and not all queries are in natural language.
Table~\ref{t02-2} lists some instances from both datasets to show their differences.

\begin{table}[t]
\caption{Some examples from the \ac{SaaC} dataset. For comparison, we also include examples from the MS MARCO dataset.}
\label{t02-2}
\begin{tabular}{@{}lm{6cm}m{7cm}@{}}
\toprule
                          & Queries                                & Responses                                                                                                                       \\ 
\midrule
\multirow{10}{*}[-3cm]{\rotatebox{90}{SaaC}}     & \raggedright What are the different forms of energy? & As far as I know, the different forms of energy are kinetic energy, potential energy, gravitational energy, nuclear energy, ... \\
                          & \raggedright How can it be stored?           & Energy can be stored in many ways like a gallon of gasoline or a barrel of oil contains stored energy that can be released when we burn it.                       \\
                          & \raggedright What type of energy is used in motion?  & The energy of motion is kinetic energy.                                       \\
                          & \raggedright Tell me about mechanical energy.              & Mechanical energy can be either kinetic energy (energy of motion) or potential energy (stored energy of position).                                                               \\ 
                          & \raggedright Give me some examples.              & Examples of mechanical energy are the power of a football flying through the air or the wrecking ball that is swung backward away from a building.                                                               \\ 
\cmidrule{2-3}                          
 & \raggedright Tell me about the Bronze Age collapse. & It may be because of a shortage of tin, that is necessary for the melt of bronze that forced to seek an alternative of this metal. \\
                          & \raggedright What is the evidence for it?           & Could be the extrinsic evidence, but I am not sure whether it can support the collapse of the Bronze Age.                       \\
                          & \raggedright What are some of the possible causes?  & One of the possible causes of the Bronze Age collapse is the invasion of the Sea Peoples.                                       \\
                          & \raggedright Who were the Sea Peoples?              & The Sea Peoples was a confederacy of seafaring raiders that caused political unrest, and attempted to enter or control Egyptian territory during the late 19th dynasty, and the 20th dynasty.                                                               \\ 
                          & \raggedright What was their role in it?              & The Sea Peoples entered and invaded eastern Mediterranean territory. The Sea People's invasions ushered or caused the bronze age collapse.                                                              \\ 
\midrule                          
\multirow{3}{*}[-0.7cm]{\rotatebox[origin=c]{90}{MS MARCO}} & \raggedright cell organelles definition                    & Cell organelles is a membrane bound compartment or structure in a cell that performs a special function.                                                                   \\
                          & \raggedright What music style developed from Scott Joplin? & Scott Joplin developed the maple leaf rag and the entertainer.                                         \\
                          & \raggedright glioma meaning                            & The meaning of glioma is a tumor springing from the neuroglia or connective tissue of the brain, spinal cord, or other portions of the nervous system.   \\ 
\bottomrule
\end{tabular}
\squeeze
\end{table}

\subsection{Description of the \ac{SaaC} dataset}
\label{section:SaaC dataset}
The \ac{SaaC} dataset has 80 topics (with a total of 748 queries, and 7--12 queries per topic) from CAsT.
Almost all of the queries come with a complex information need; 359 are ``what'' queries, 144 are ``how'' queries, and 47 are ``why'' queries.
It is important to note that, compared with CAsT, \ac{SaaC} has three types of additional annotations: the reformulated queries, the supporting spans from the relevant passages, and the conversational responses.

We report some descriptive statistics for the \ac{SaaC} dataset in Table~\ref{t04-1}; for comparison we include the same information for the MS MARCO dataset.
\begin{table}[t]
\centering
\caption{Descriptive statistics for the \ac{SaaC} dataset. For comparison, we include the same information for the  MS MARCO dataset.}\label{t04-1}
\begin{tabular}{lcc}
                             \toprule
                             & SaaC & MS MARCO    \\
                             \midrule
\#query length                & \phantom{0}7.21\phantom{\%}    & \phantom{0}6.08\phantom{\%}     \\
\#answer length              & 28.19\phantom{\%} & 15.90\phantom{\%}       \\
\#passage length            & \llap{1}55.25\phantom{\%}   & 67.11\phantom{\%}     \\
\#pairwise passage similarity  & 31.14\% & 26.77\%  \\
\#1-gram overlap             & 80.75\% & 90.93\%   \\
\#2-gram overlap            & 47.68\%  & 72.92\%   \\
\#3-gram overlap             & 34.85\%  & 63.43\%  \\
\#4-gram overlap             & 27.85\%  & 58.01\%  \\
\#query common words ratio   & 70.80\%  & 56.15\%  \\
\#answer common words ratio  & 60.21\% & 53.61\%   \\
\bottomrule
\end{tabular}
\squeeze
\end{table}
In the table,
``\#pairwise passage similarity'' denotes the average TF-IDF dot similarity of each pair of candidate passages.
``\#n-gram overlap'' denotes the average n-gram overlap ratio between the answer and candidate passages.
``query/answer common words ratio'' denotes the ratio of common words (word frequency $\geq$100,000) in the query and answer, respectively.
We can see from Table~\ref{t04-1} that \ac{SaaC} is more complex and conversational than MS MARCO.
\ac{SaaC} is more complex in that
\begin{enumerate*}
\item the ``\#query/answer/passage length'' is larger, which means the queries are more complex to understand, and the passages contain more (noisy) information; and
\item the ``\#pairwise passage similarity'' is larger, which means it has more confusing candidate passages, making it hard to find the correct ones.
\end{enumerate*}
\ac{SaaC} is also more conversational because:
\begin{enumerate*}
\item the ``n-gram overlap'' between the answers and passages is much smaller, which means the answers are more abstractive; and
\item the ``\#query/answer common words ratio'' is larger, which means that the queries and answers are more in spoken language, which is more informal.
\end{enumerate*}

%% file: sections/03-model.tex

\section{A Method for Conversing with Search Engines}
\label{03-model}

\begin{table*}[]
\caption{Summary of main symbols and notation used in the paper.}
\label{t03-1}
\begin{tabular}{lm{12cm}}
\toprule
$q$                 & A user query. If there are superscript and/or subscript, superscript means the conversation turn and subscript means the token index, e.g, $q^t_i$.                                                 \\
$Q^t$               & A list of queries until $t$th turn.                                                                                                                                                                 \\
$d$                 & A passage. If there are superscript and/or subscript, superscript means the passage index in the candidate set and subscript means the token index, e.g, $d^k_i$.                                   \\
$D^t$               & The passage candidate list for the $t$th turn.                                                                                                                                                      \\
$\mathbf{h}$        & Hidden representation for a token. The superscript and/or subscript indicate the definition (what it models), the type (query or document), the token index and so on, e.g, $\mathbf{h}_{q^{t}_i}$. \\
$H$                 & A sequence of representations.                                                                                                                                                                      \\
$r$                 & A response. If there are superscript and/or subscript, superscript means the conversation turn and subscript means the token index, e.g, $r^t_j$.                                                   \\
$P()$               & Probability.                                                                                                                                                                                        \\
$W$ or $\mathbf{v}$ & Learnable parameters.                                                                                                                                                                               \\
$\hat{y}$ or $y$    & (Weak) ground truth label. \\
\bottomrule
\end{tabular}
\end{table*}

Users could have conversations with search engines for many reasons, like getting information, performing tasks, etc. 
In this work, we only focus on users' information exploration needs, in a setting where users express their needs in natural language and the system returns the information the users need in short conversational responses.

Table~\ref{t03-1} summarizes the  main symbols and notation used in this paper.
Formally, given a series of previous user queries $Q^{t-1}=[q^{t-m}$, $q^{t-m+1}$, $\ldots,$ $q^{t-1}]$ in natural language, the user query at the current turn $q^t$, and a list of candidate passages $D^t=[d^1$, $d^2$, $\ldots,$ $d^k]$ that have been retrieved by a search engine and potentially contain the answers (or are at least relevant to $q^t$),\footnote{While creating effective queries from multi-turn questions is a challenging task in itself, we assume that the candidate documents are provided in advance so as to simplify the experimental design and facilitate reproducibility.} the \ac{SaaC} task is to generate a short response $r^t$ for $q^t$ by identifying the relevant passages and summarizing the supporting spans from these passages into a conversational response.
Note that we do not take the previous system responses as input, because: 
\begin{enumerate*}
\item this might result in error accumulation when the previous system responses are improper in practice; and 
\item in our datasets, all the background information that is needed to understand the current query is covered by the previous user queries.
\end{enumerate*}
But, it is easy to extend all the models including the baselines and ours to consider the previous system responses.

\begin{figure*}[t]
\centering
\includegraphics[clip,trim=-2mm 0mm 0mm 0mm,width=1\textwidth]{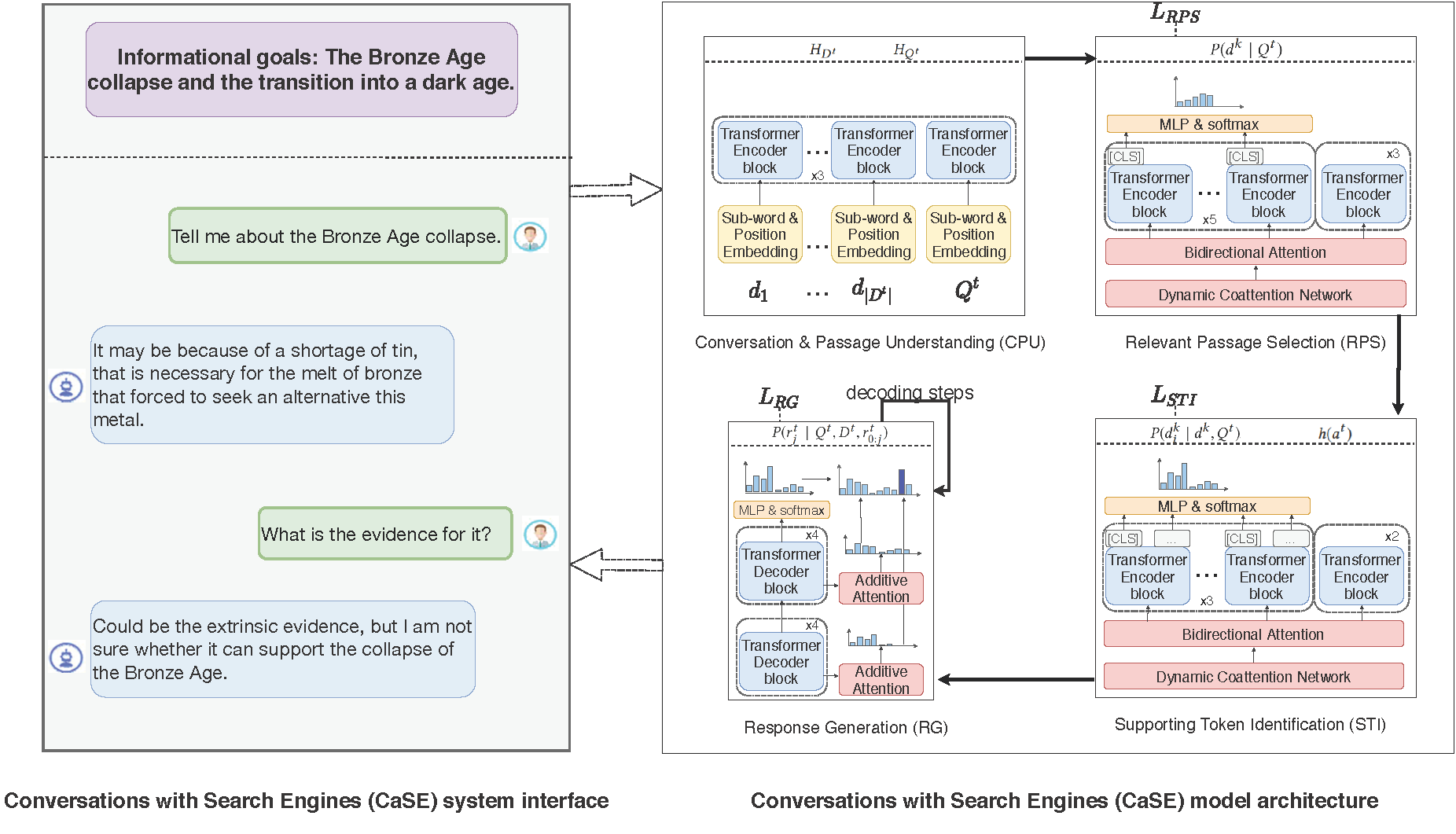}
\caption{An overview of \acf{CaSE}. Section~\ref{03-model} contains a walkthrough of the model.}
\label{03-1}
\end{figure*}

In this work, we decompose the \ac{SaaC} task into four sub-tasks: 
\begin{enumerate*}
\item \acf{CPU}, 
\item \acf{RPS}, 
\item \acf{STI}, and 
\item \acf{RG}, 
\end{enumerate*}
as shown in Figure~\ref{03-1}.
We devise a modularized framework, \acf{CaSE}, to operationalize the sub-tasks in an end-to-end manner.
Both MS MARCO and our \ac{SaaC} still adopt a ``user asks -- system responds'' paradigm which does not consider mixed-initiative~\cite{Hamedwww2020}, so we preclude this part in the \ac{CaSE} framework.

Specifically, the \ac{CPU} module first encodes each query $q$ and passage $d$ into a sequence of hidden representations, i.e., 
\begin{align}
H_{q^t}&=[\mathbf{h}_{q^{t}_1}, \ldots, \mathbf{h}_{q^{t}_{|q^t|}}]\text{ for }q^t\text{, and}
\label{eq:Hqt}\\
H_{d^k}&=[\mathbf{h}_{d^{k}_1}, \ldots, \mathbf{h}_{d^{k}_{|d^{k}|}}]\text{ for }d^k.
\label{eq:Hdk}
\end{align}
Then, based on the query and passage representations, the \ac{RPS} module selects the relevant passage by estimating the passage relevance probability $P(d^k \mid Q^t)$ for each passage $d^k$ in the candidate pool $D^t$.
After that, the \ac{STI} module estimates the probability of each passage token to be a supporting token, $P(d^k_i \mid d^k, Q^t)$, where a token is supporting if it contributes to the final responses.
Finally, the \ac{RG} module takes the outputs from the above three modules into consideration and generates a short response, token by token.
In particular, $P(d^k \mid Q^t)$ and $P(d^k_i \mid d^k, Q^t)$ are modeled as priors in the \ac{RG} module.

In the following subsections, we will describe our proposed solution for each module in detail.

\subsection{Conversation and Passage Understanding}
\label{s03-1}

We employ a Transformer model to perform conversation and passage understanding, which relies on self-attention to extract important information to represent conversations and passages.
Specifically, for the conversational queries $Q^{t}=Q^{t-1}\cup \{q^t\}$, we first concatenate the tokens as one sequence, and then input it into a stack of Transformer encoder blocks~\citep{10.5555/3295222.3295349} to obtain representations for each query and each token $H_{Q^{t}}=[H_{q^{t-m}}$, $H_{q^{t-m+1}}$, $\ldots,$ $H_{q^{t}}]$, where each $H_{q^t}$ is defined as in Eq.~\ref{eq:Hqt}.
Note that we put a special token ``[CLS]'' at the start, formally referred to as $\mathbf{h}_{q^{t}_{CLS}}$, which is considered to represent the conversations up to the current conversation turn $t$.
Similarly, we obtain representations $H_{d^k}$ as in Eq.~\ref{eq:Hdk}
for each passage.

\subsection{Relevant Passage Selection}
\label{s03-2}

In order to model the relevance to conversational queries of each passage, we first need to model the interaction between them.
Here, we employ a similar bi-directional attention flow as proposed by~\citet{Seo2017Bidirectional} to do \acf{MRC}, which is also used by~\citet{nishida-etal-2019-multi} to do \acf{QA}.
Specifically, we first obtain the interaction matrix $M^{Qd_k} \in$ $\mathbb{R}^{|Q_t| \times |d_k| \times N}$ between the conversation tokens $H_{Q^{t}}$ and each passage $H_{d^k}$~\cite{Seo2017Bidirectional} as follows:
\begin{equation}
\label{equation_ps_1}
M^{Qd_k} = 
\begin{bmatrix}
  f^{Qd_k}(\mathbf{h}_{q^{t-m}_1}, \mathbf{h}_{d^k_1}) & f^{Qd_k}(\mathbf{h}_{q^{t-m}_1}, \mathbf{h}_{d^k_2}) & \cdots & f^{Qd_k}(\mathbf{h}_{q^{t-m}_1}, \mathbf{h}_{d_{|d^k|}}) \\
  f^{Qd_k}(\mathbf{h}_{q^{t-m}_2}, \mathbf{h}_{d^k_1}) & f^{Qd_k}(\mathbf{h}_{q^{t-m}_2}, \mathbf{h}_{d^k_2}) & \cdots & f^{Qd_k}(\mathbf{h}_{q^{t-m}_2}, \mathbf{h}_{d_{|d^k|}}) \\
  \cdots & \cdots & \ddots & \cdots \\
  f^{Qd_k}(\mathbf{h}_{q^{t}_{|q^{t}|}}, \mathbf{h}_{d^k_1}) & f^{Qd_k}(\mathbf{h}_{q^{t}_{|q^{t}|}}, \mathbf{h}_{d^k_2}) & \cdots & f^{Qd_k}(\mathbf{h}_{q^{t}_{|q^{t}|}}, \mathbf{h}_{d^k_{|d^k|}})
\end{bmatrix}
,
\end{equation}
where $f^{Qd_k}$ is the cross-correlation function, which captures the correlation between each pair of tokens in queries and passages:
\begin{equation}
\label{equation_ps_2}
f^{Qd_k}(\mathbf{h}_{q_i}, \mathbf{h}_{d^k_j}) = {\mathbf{v}^{Qd_k}}^\top [\mathbf{h}_{q_i} \oplus \mathbf{h}_{d^k_j} \oplus (\mathbf{h}_{q_i} \odot \mathbf{h}_{d^k_j})],
\end{equation}
where ${\mathbf{v}^{Qd_k}} \in \mathbb{R}^{3N \times 1}$ is the parameter vector; $\oplus$ denotes the concatenation operation and $\odot$ denotes the Hadamard product.
Then, a Dynamic Coattention Network as proposed in~\cite{Xiong2017fast} is used to obtain the dual attention representations for the conversational queries $H_{Q^t \leftharpoonup D^t}$, and each passage $H_{d^k \leftharpoonup Q^t}$ as follows:
\begin{align}
\label{equation_ps_3}
\begin{split}
H_{Q^t \leftharpoonup D^t} = {} & [H_{Q^t} \oplus H^{1}_{Q^t \leftharpoonup D^t} \oplus H^{2}_{Q^t \leftharpoonup D^t} \oplus (H^{1}_{Q^t \leftharpoonup D^t} \odot H_{Q^t}) \oplus (H^{2}_{Q^t \leftharpoonup D^t} \odot H_{Q^t})] \\
H_{d^k \leftharpoonup Q^t} = {} & [H_{d^k} \oplus H^{1}_{d^k \leftharpoonup Q^t} \oplus H^{2}_{d^k \leftharpoonup Q^t} \oplus (H^{1}_{d^k \leftharpoonup Q^t} \odot H_{d^k}) \oplus (H^{2}_{d^k \leftharpoonup Q^t} \odot H_{d^k})].
\end{split}
\end{align}
Take $H_{Q^t \leftharpoonup D^t}$, for instance.
It contains five parts: $H_{Q^t}$ is the preserved information from the queries;
$H^{1}_{Q^t \leftharpoonup D^t}$ and $H^{2}_{Q^t \leftharpoonup D^t}$ represent the preserved information from the passages after the first- and second-order selection with the help of the interaction matrix $M^{Qd_k}$;
$(H^{1}_{Q^t \leftharpoonup D^t} \odot H_{Q^t})$ and $(H^{2}_{Q^t \leftharpoonup D^t} \odot H_{Q^t})$ capture the further interaction between two types of information.
The same is true for $H_{d^k \leftharpoonup Q^t}$.
Here, $H^{1}_{Q^t \leftharpoonup D^t}$, $H^{2}_{Q^t \leftharpoonup D^t}$, $H^{1}_{d^k \leftharpoonup Q^t}$ and $H^{2}_{d^k \leftharpoonup Q^t}$ are defined as follows:
\begin{align}
H^{1}_{Q^t \leftharpoonup D^t}&=\max (\{M_{d_k} H_{d^k}\}_{k=1,\ldots, |D^t|}) 
  \label{eq:first}\\
H^{2}_{Q^t \leftharpoonup D^t}&=\max (\{M_{d^k} H^{1}_{d^k \leftharpoonup Q^t}\}_{k=1,\ldots, |D^t|})
  \label{eq:second}\\
H^{1}_{d^k \leftharpoonup Q^t}&=M_Q^\top H_{Q^t} 
  \label{eq:third}\\
H^{2}_{d^k \leftharpoonup Q^t}&=M_Q^\top H^{1}_{Q^t \leftharpoonup D^t} 
  \label{eq:fourth}\\
M_{d_k}&=\softmax_{d}(M^{Qd_k}) 
  \label{eq:fifth}\\
M_Q&=\softmax_{Q}(M^{Qd_k}).
  \label{eq:sixth}
\end{align}
In Equation~\ref{eq:first}--\ref{eq:sixth}, $\max$ denotes max pooling; $\softmax_{Q}$ and $\softmax_{d}$ denote the softmax over $M^{Qd_k}$ along the query and passage dimension, respectively.
Then, a stack of the Transformer encoder blocks are used to reduce the dimension of $H_{Q^t \leftharpoonup D^t}$ and $H_{d^k \leftharpoonup Q^t}$.
By conducting the above operations, the model can learn to control the relevant information flow from queries to passages (e.g., $D^t \leftharpoonup Q^t$, $d^k \leftharpoonup Q^t$), and vice versa.

To estimate the passage relevance score, we use an MLP to get the passage relevance score by taking the first token representation of each passage $H_{d^k_0 \leftharpoonup Q^t}$ (corresponding to the ``[CLS]'' token in Section~\ref{s03-1}) as input.
The relevance score is normalized with a sigmoid to obtain the relevance probability $P(d^k \mid Q^t)$.
We use binary cross entropy to supervise the learning of this module:
\begin{equation}
\label{equation_ps_4}
\begin{split}
L&_{RPS}  = {} \sum_{d^k \in D^t} y_{d^k} \log P(d^k \mid Q^t) + (1-y_{d^k}) \log (1-P(d^k \mid Q^t)),
\end{split}
\end{equation}
where $y_{d^k}=1$ if $d$ is, relevant otherwise $y_{d^k}=0$.

\subsection{Supporting Token Identification}
Previous methods directly generate the responses after passage selection, which we hypothesize can be improved by incorporating a dedicated \acf{STI} module.
The core idea is that, besides learning to select the relevant passage, the model could also learn to identify supporting tokens, which might be used to generate the response.

To do so, we use a similar architecture as in the \ac{RPS} module (parameters are not shared) to get updated representations $H_{Q^t \leftharpoonup D^t}$ and $H_{D^t \leftharpoonup Q^t}$.
But, instead of estimating the passage relevance score based on the ``[CLS]'' representation (which represents the general representation of the passage), we estimate the probability of each passage token as a supporting token $P(d^k_i \mid d^k, Q^t)$.
Specifically, we use an MLP to get the supporting token likelihood score for each token with \smash{$h_{d^k_i \leftharpoonup Q^t} \in H_{D^t \leftharpoonup Q^t}$} as input ($h_{d^k_i \leftharpoonup Q^t}$ represents the token representation for $d^k_i$), which is normalized with a sigmoid to obtain $P(d^k_i \mid d^k, Q^t)$.

Unfortunately, there are no ground truth labels to define a supervised learning loss to train $P(d^k_i \mid d^k, Q^t)$.
To this end, we design a weak supervision signal based on the following intuitions: 
\begin{enumerate*}
\item if a passage token $d^k_i$ is a supporting token, it must exist in the ground truth response;
\item if the surrounding tokens of $d^k_i$ are also in the ground truth response, $d^k_i$ is more likely to be a supporting token; and
\item rare passage tokens that exist in the ground truth response are more likely to be supporting tokens than frequent ones.
\end{enumerate*}
Specifically, we devise a \acfi{CCCE} loss as follows:
\begin{equation}
\label{equation_se_4}
\begin{split}
L_{STI} = -\sum_{d^k \in D^t} \sum_{d^k_i \in d^k} & \left[ c(d^k_i) \hat{y}_{d^k_i}\log P(d^k_i \mid d^k, Q^t) + (1-\hat{y}_{d^k_i})\log (1-P(d^k_i \mid d^k, Q^t))\right],
\end{split}
\end{equation}
where \smash{$\hat{y}_{d^k_i}$} is a weak label indicating whether \smash{$d^k_i$} is a supporting token; \smash{$\hat{y}_{d^k_i}=1$} if \smash{$d^k_i \in r^{t*}$}, and 0 otherwise, where $r^{t*}$ is the ground truth response;
and \smash{$c(d^k_i)$} is a coefficient indicating the confidence of $d^k_i$ as a supporting token, which is defined as:
\begin{equation}
\label{equation_se_5}
c(d^k_i) \propto \frac{1}{\log \mathit{freq}(d^k_i)} \cdot \prod_{n} |d^k_{i-n:i+n} \cap r^{t*}|,
\end{equation}
where \smash{$\mathit{freq}(d^k_i)$} is the token frequency in the data collection.
The first term models how ``rare'' \smash{$d^k_i$} is, the second term models how likely \smash{$d^k_i$} and its $n$ surrounding tokens are supporting tokens (overlapping with the ground truth).
Finally, \smash{$c(d^k_i)$} ensures that rare and more often overlapping tokens get more opportunities to be identified as supporting tokens.

\subsection{Response Generation}
We propose a \acf{PPG} to implement the \ac{RG} module, which is able to generate tokens from a predefined vocabulary and copy tokens from both the queries and passages.
The idea is that each token can be generated under three modes, i.e., with probability $P(r^t_j \mid Q^t, D^t, r^t_{0:j}, g)$ from the generation mode $g$ (a.k.a. the \textit{vocabulary generator} which generates from a predefined vocabulary), with probability $P(r^t_j \mid Q^t, D^t, r^t_{0:j}, c_{Q^t})$ from the query copying mode $c_{Q^t}$ (a.k.a. the \textit{query pointer generator} which copies from queries), and with probability $P(r^t_j \mid Q^t, D^t, r^t_{0:j}, c_{D^t})$ from the passage copying mode $c_{D^t}$ (a.k.a. the \textit{prior-aware passage pointer generator} which copies from passages).
The final probability of generating the token $r^t_j$ is a combination of the three probabilities.
Especially when estimating the passage copying probability, \ac{PPG} models the passage relevance and supporting token likelihood from the \ac{RPS} and \ac{STI} modules as priors.

Given the previous decoded tokens $r^t_{0:j}=[r^t_{0}, \ldots, r^t_{j-1}]$ ($r^t_{0}$ is set to a special token ``[BOS]'' indicating the beginning of decoding), we first use a stack of Transformer decoder blocks~\cite{10.5555/3295222.3295349} to obtain the hidden representations 
\[
H^{Q}_{r^t_{0:j}} = [h^{Q}_{r^t_{0}}, \ldots, h^{Q}_{r^t_{j-1}}],
\]
which takes \smash{$r^t_{0:j}$} and \smash{$H_{Q^t \leftharpoonup D^t}$} as inputs.
Then, we use the same stack of Transformer Decoder blocks to obtain hidden representations 
\begin{equation}
H^{D}_{r^t_{0:j}} = [h^{D}_{r^t_{0}}, \ldots, h^{D}_{r^t_{j-1}}],
\end{equation}
which takes \smash{$H^{Q}_{r^t_{0:j}}$} and \smash{$H_{D^t \leftharpoonup Q^t}$} as inputs.
Afterwards, we estimate the token probability from three modes: generating from the vocabulary $g$, copying from queries $c_{Q^t}$, and copying from passages $c_{D^t}$.

\paragraph{Vocabulary generator} The probability of generating a token from the predefined vocabulary is estimated as:
\begin{equation}
\label{equation_rg_1}
\begin{split}
P(r^t_j \mid Q^t, {}&{} D^t, r^t_{0:j}, g) = {} P(g \mid r^t_{0:j}) \softmax(\mlp([r^t_{j-1} \oplus h^{D}_{r^t_{j-1}} \oplus h(a^t)])),
\end{split}
\end{equation}
where \smash{$P(g \mid r^t_{0:j})$} denotes the probability of the generation mode $g$; \smash{$h(a^t)$} is the answer representation from the \ac{STI} module, which is estimated as follows:
\begin{equation}
h(a^t) = \sum_{d^k \in D^t}{P(d^k \mid Q^t) \sum_{d^k_i \in d^k}{P(d^k_i \mid d^k, Q^t)h_{d^k_i \leftharpoonup Q^t}}},
\end{equation}
where \smash{$P(d^k \mid Q^t)$} is the passage relevance probability; \smash{$h_{d^k_i \leftharpoonup Q^t}$} is the $i$-th token representation from \smash{$H_{d^k \leftharpoonup Q^t}$}.
Both are from the \ac{RPS} module.
$P(d^k_i \mid d^k, Q^t)$ is the supporting token probability from the \ac{STI} module.

\paragraph{Query pointer generator}
We use another additive attention to estimate the probability of copying a token from the conversational queries:
\begin{equation}
\begin{split}
P(r^t_j=Q^t_i \mid{}& Q^t, D^t, r^t_{0:j}, c_{Q^t})={} P(c_{Q^t} \mid r^t_{0:j}) P(Q^t_i \mid Q^t, D^t, r^t_{0:j}),
\end{split}
\end{equation}
where $P(c_{Q^t} \mid r^t_{0:j})$ is the query copying mode probability; and $P(Q^t_i \mid Q^t, D^t, r^t_{0:j})=\attention(query: h^{Q}_{r^t_{j-1}}, key: H_{Q^t \leftharpoonup D^t})$.

\paragraph{Prior-aware passage pointer generator} 
The probability of copying a token from the passages is
\begin{equation}
\label{equation_rg_2}
\begin{split}
P&(r^t_j=d^k_i \mid Q^t, D^t, r^t_{0:j}, c_{D^t}) \\
& =   P(c_{D^t} \mid r^t_{0:j}) \hspace*{-1mm}\sum_{d^k \in D^t} P(d^k \mid Q^t)\cdot\sum_{d^k_i \in d} P(d^k_i \mid d^k, Q^t) P(d^k_i \mid Q^t, D^t, r^t_{0:j}),
 \end{split}
\end{equation}
where \smash{$P(c_{D^t} \mid r^t_{0:j})$} is the passage copying mode probability; {$P(d^k \mid Q^t)$} is the passage prior from the \ac{RPS} module and $P(d^k_i \mid d^k, Q^t)$ is the supporting token prior from the \ac{STI} module; and $P(d^k_i \mid Q^t, D^t, r^t_{0:j})=$ $\attention(query: h^{D}_{r^t_{j-1}}, key: H_{D^t \leftharpoonup Q^t})$.
Here, the prior-aware passage pointer generator learns to copy tokens from all the candidate passages despite that we can predict which passage is most relevant from the \ac{RPS} module.
The reason is that it is hard for the \ac{RPS} module to always rank the most relevant passage as top one.


To coordinate the probabilities from different modes, we learn a mode coordination probability:
\begin{equation}
\label{equation_rg_3}
\begin{split}
[P(g \mid r^t_{0:j}), P(c_{Q^t} \mid r^t_{0:j}), & P(c_{D^t} \mid r^t_{0:j})] = {} W^\top [h_{r^t_{j-1}} \oplus h^{att}_{Q^t} \oplus h^{att}_{D^t}],
\end{split}
\end{equation}
where $W \in 3N \times 3$ is the parameter matrix;
\smash{$h^{att}_{Q^t}$} and \smash{$h^{att}_{D^t}$} are the attended query and passage representations from the two attentions in the query and passage pointer generators, respectively.
The final probability at the $j$-th decoding step is
\begin{equation}
P(r^t_j \mid Q^t, D^t, r^t_{0:j}) =P(r^t_j \mid Q^t, D^t, r^t_{0:j}, g) +P(r^t_j \mid Q^t, D^t, r^t_{0:j}, c_{Q^t})+ P(r^t_j \mid Q^t, D^t, r^t_{0:j}, c_{D^t}).
\end{equation}
If a token is absent from a mode, its corresponding probability from that mode is set to zero.

We use the negative log likelihood loss to train \ac{PPG} as follows:
\begin{equation}
\label{equation_rg_5}
L_{RG} = -\sum_{r^t_j \in r^t} \log P(r^t_j \mid Q^t, D^t, r^t_{0:j}).
\end{equation}

%% file: sections/04-experimentsetup.tex

\section{Experimental Setup}
\label{section:ExperimentalSetup}

We seek to answer the following research questions by designing experiments:
\begin{enumerate}[leftmargin=*,label=(Q\arabic*)]
\item What is the performance of \ac{CaSE} compared to other methods? Does \ac{CaSE} outperform the state-of-the-art methods in terms of response generation and passage ranking performance? What, if any, are the performance differences on the MS MARCO and \ac{SaaC} datasets?
\item What is the effect of different components of \ac{CaSE} on its overall performance?
\item Where does \ac{CaSE} fail? That is, is there any room for further improvement on the \ac{SaaC} dataset? 
\end{enumerate}

\subsection{Datasets and Evaluation Metrics}

As pointed out in Section~\ref{section:SaaC dataset}, \ac{SaaC} contains 80 topics (with 748 queries) from CAsT, which is too small to train neural models in an end-to-end fashion.
Hence, we train all the models that we consider on the MS MARCO 2.1 Q\&A + Natural Language Generation training set (MS MARCO train)~\cite{DBLP:conf/nips/NguyenRSGTMD16}.
Although this is a single-turn dataset and the queries and responses are less conversational than in the \ac{SaaC} dataset (see Table~\ref{t02-2}), it does have passage relevance labels and human written answers, which is sufficient as a training set.
MS MARCO is the sole dataset with real queries from search engines and human written answers.
There are two types of answer in the MS MARCO dataset: QA style (corresponding to the ``answers'' field in the dataset) and NLG style (corresponding to the ``wellFormedAnswers'' field in the dataset). 
The QA style answers are not always in natural language, according to the official description.\footnote{\url{https://github.com/microsoft/MSMARCO-Question-Answering}}
The NLG style answers were generated by rewriting the QA style answers to make sure that:
\begin{enumerate*}
\item each NLG style answer includes proper grammar to make it a full sentence; 
\item the NLG style answers make sense without the context of either the query or the passage; and
\item the NLG style answers have a high overlap with exact portions in one of the context passages. 
\end{enumerate*}
This ensures that NLG style answers are true natural language and not just selected spans.
In this work, we focus on generating natural language responses, so we only keep data samples where the ``wellFormedAnswers'' field is not empty.
We randomly split the original development set into two sets with roughly equal size, one as our development set (MS MARCO dev) and the other as our test set (MS MARCO test).
The sizes of the MS MARCO train, MS MARCO dev, MS MARCO test, and \ac{SaaC} test are 524,105, 32,345, 32,225, and 1,008,\footnote{Some queries have more than one relevant passage. We construct a test sample for each relevant passage.} respectively.
Although there are other multi-turn conversational datasets, e.g., those shown in Table~\ref{t02-1}, we do not report results on them as they are not for the search scenario.

We use BLEU (up to 4-grams using uniform weights), and ROUGE-1, ROUGE-2, and ROUGE-L to evaluate the response generation performance, which are commonly used in natural language generation tasks, e.g., \ac{QA}, \ac{MRC}~\citep{kocisky-etal-2018-narrativeqa,nishida-etal-2019-multi}.
We also report MAP, Recall@5 and NDCG for passage ranking performance.





\subsection{Methods Used for Comparison}


We collect and implement state-of-the-art methods from various related tasks.

\begin{itemize}
\item \textbf{S2SA}. S2SA is a sequence-to-sequence with attention model; it is commonly used as a baseline for natural language generation tasks~\cite{moghe2018towards,meng-2020-refnet}.
\item \textbf{GTTP}~\cite{see-etal-2017-get}. GTTP improves S2SA by incorporating a pointer mechanism that enables it to copy tokens from the input during generation. GTTP achieves state-of-the-art performance on many natural language generation tasks~\cite{paulus2017deep,reddy-etal-2019-coqa}.
\item \textbf{TMemNet}~\cite{dinan2018wizard}. TMemNet was first introduced for the knowledge grounded dialogue task. It combines a Memory Network and Transformer to do knowledge selection and dialogue response generation.
\item \textbf{GLKS}~\cite{ren-2020-thinking}. GLKS is a state-of-the-art method for \ac{BBC} (the best performing method on the Holl-E\footnote{\url{https://github.com/nikitacs16/Holl-E}} dataset at the time of writing). It introduces a mechanism to combine global and local knowledge selection for dialogue response generation.
\item \textbf{Masque}~\cite{nishida-etal-2019-multi}. Masque is the best performing method on the MS MARCO Q\&A + Natural Language Generation task at the time of writing. Because we only use ``wellFormedAnswers'' to train the models, we remove the answer possibility classifier and style token.
\item \textbf{\ac{CaSE}}. \ac{CaSE} is proposed in this paper.
\end{itemize}

\subsection{Implementation Details}
For a fair comparison, we implement all models used in our experiments based on the same code framework to ensure that they share the same code apart from the model itself.
We set the word embedding size and hidden size to 256.
We use the BERT vocabulary\footnote{\url{https://github.com/huggingface/transformers}} for all methods but for a fair comparison, we avoid using any extra resources for all methods, including pre-trained embeddings.
The vocabulary size is 30,522.
The learning rate was increased linearly from zero to $2.5\times10^{-4}$ in the first 6,000 steps and then annealed to 0 by using a cosine schedule. 
We use gradient clipping with a maximum gradient norm of 1.
We use the Adam optimizer ($\alpha = 0.001$, $\beta_1 = 0.9$, $\beta_2 = 0.999$, and $\epsilon$ = $10^{-8}$).
An exponential moving average was applied to all trainable variables with a decay rate of 0.9995.
For all models, we combine multiple losses linearly if there is more than one.
For TMemNet, Masque and CaSE, the parameters of the Transformer encoder for the queries and passages are shared, as we found that this resulted in better performance.
We train all models on four TITAN X (Pascal) GPUs.
The batch size is chosen from (32, 64, 128) according to the GPU memory.
We select the best models based on performance on the development set.

%% file: sections/05-results.tex

\section{Experimental results}
\label{section:ExperimentalResults}

\subsection{How Does \ac{CaSE} Perform?}
\label{s05-1}

To answer Q1, we report the results of all methods on both MS MARCO (see Table~\ref{t05-1-1}) and \ac{SaaC} (see Table~\ref{t05-1-2}).
S2SA, GTTP and GLKS do not perform passage ranking, so there are no MAP, Recall@5 and NDCG results for those methods.
Given the results, there are three main lessons.

\begin{table*}[t]
\caption{Overall performance on the MS MARCO test set (\%). \textbf{Bold face} indicates the best result in terms of the corresponding metric. R-1: ROUGE-1; R-2: ROUGE-2; R-L: ROUGE-L. $^*$ and $^{**}$ indicate \ac{CaSE} is significantly better than Masque (p-value < 0.05 and p-value < 0.01 with t-test, respectively).
Note that the scores are not comparable to the leaderboard of MS MARCO at \url{http://www.msmarco.org/leaders.aspx} as different training and test sets are used.
}
\label{t05-1-1}
\begin{tabular}{lccccccc}
\toprule
\multirow{2}{*}{Methods} & \multicolumn{7}{c}{MS MARCO test}                                                                                             \\
\cmidrule(l){2-8} 
                         & R-1            & R-2            & R-L            & BLEU           & MAP            & Recall@5        & NDCG                    \\
                         \midrule
S2SA                     & 46.75          & 25.17          & 38.79          & 18.61          & --              & --              & --             \\
GTTP                     & 47.41          & 25.74          & 39.64          & 19.32          & --              & --              & --  \\
GLKS                     & 47.74          & 27.51          & 40.01          & 20.75          & --              & --              & --   \\
TMemNet       & 48.01          & 29.56          & 40.29          & 23.23          & 53.43          & 82.56          & 64.83         \\
Masque                   & 55.33          & 37.15          & 47.75          & 30.01          & 63.75          & 90.77          & 72.82  \\
CaSE                      & \textbf{57.44}\rlap{$^{**}$} & \textbf{39.26}\rlap{$^{**}$} & \textbf{49.91}\rlap{$^{**}$} & \textbf{32.08}\rlap{$^{**}$} & \textbf{65.75}\rlap{$^{**}$} & \textbf{92.20}\rlap{$^{**}$} & \textbf{74.34}\rlap{$^{**}$} \\ 
\bottomrule
\end{tabular}
\end{table*}

\begin{table*}[t]
\caption{Overall performance on the SaaC test set (\%). \textbf{Bold face} indicates the best result in terms of the corresponding metric. R-1: ROUGE-1; R-2: ROUGE-2; R-L: ROUGE-L. $^*$ and $^{**}$ indicate \ac{CaSE} is significantly better than Masque (p-value < 0.05 and p-value < 0.01 with t-test, respectively).}
\label{t05-1-2}
\begin{tabular}{lccccccc}
\toprule
\multirow{2}{*}{Methods} & \multicolumn{7}{c}{SaaC test}                                                                                        \\
\cmidrule(l){2-8} 
                         & R-1            & R-2            & R-L            & BLEU           & MAP            & Recall@5        & NDCG            \\
                         \midrule
S2SA                               & 24.71          & 09.04          & 17.69          & 07.23          & --              & --              & --              \\
GTTP                                 & 28.89          & 10.51          & 20.61          & 08.80          & --              & --              & --              \\
GLKS                                  & 35.46          & 13.46          & 25.31          & 12.38          & --              & --              & --              \\
TMemNet       & 35.29          & 12.72          & 25.05          & 09.60          & 14.12          & 11.31          & 29.25          \\
Masque                   & 35.44          & 13.48          & 25.74          & 11.97          & 16.13          & 12.00          & 31.77          \\
CaSE                      & \textbf{37.34}\rlap{$^*$} & \textbf{14.58} & \textbf{27.24}\rlap{$^*$} & \textbf{13.26} & \textbf{17.22} & \textbf{13.72} & \textbf{32.67} \\ 
\bottomrule
\end{tabular}
\end{table*}

First, \ac{CaSE} achieves the best results on both datasets in terms of all metrics.
Especially, \ac{CaSE} outperforms the best performing model Masque on the MS MARCO leaderboard and the best performing model GLKS from the \ac{BBC} task (at the time of writing).
Generally, \ac{CaSE} improves over Masque by around 2\%pt in terms of generation evaluation metrics and around 1--2\%pt in terms of passage ranking metrics.
Part of the improvement is from the proposed \ac{STI} and \ac{PPG} modules, which we will analyze in more detail in \S\ref{s05-2}.

Second, the scores on the MS MARCO dataset are much higher than those on the \ac{SaaC} dataset. 
This is the case for all methods, including \ac{CaSE}.
For example, the BLEU score of \ac{CaSE} is 18.82\%pt higher on MS MARCO for response generation, and the MAP score is 48.53\%pt higher.
This confirms that \ac{SaaC} is a more challenging and more suitable dataset than MS MARCO for research on conversations with search engines.
\ac{SaaC} is more challenging because 
\begin{enumerate*}
\item the queries and passages are more complex, which is clear from Table~\ref{t04-1}.
\item \ac{SaaC} has greater query, passage and answer lengths and the passages are more similar.
\end{enumerate*}
\ac{SaaC} is more suitable for conversations with search engines because 
\begin{enumerate*}
\item it has multi-turn conversations, which introduces the requirements of modeling context from historical turns. 
\item the responses are more abstractive and conversational, which is closer to real conversation scenarios.
\end{enumerate*}
See Section~\ref{s05-3} for further details.

Third, modeling passage selection is necessary. 
On the one hand, we can see that the methods with passage selection modules (TMemNet, Masque and \ac{CaSE}) are generally much better than those without (i.e., S2SA, GTTP and GLKS).
On the other hand, we also notice that the improvements for passage ranking are consistent with the improvements for response generation.

\subsection{What Do the Components of \ac{CaSE} Contribute?}
\label{s05-2}
To answer Q2 and analyze the effects of the \ac{RPS}, \ac{STI} and \ac{RG} modules in \ac{CaSE}, we conduct an ablation study.
We do not separately study the modeling and learning of the \ac{STI} module, as removing the $L_{STI}$ loss will result in a lack of direct supervision to guide the learning of \ac{STI}, which does not make sense.
The results on the MS MARCO and SaaC test sets are shown in Table~\ref{t05-2-1} and \ref{t05-2-2}, respectively.

\begin{table*}[t]
\caption{Ablation study on the MS MARCO test set (\%).
CaSE-X denotes CaSE with the component X left out.
CaSE-RG: replacing the \ac{RG} component with a traditional pointer generator.
CaSE-STI: \ac{CaSE} without the \ac{STI} component.
CaSE-RPS: \ac{CaSE} without the \ac{RPS} module. 
There is no CaSE-CPU because all other components rely on the \ac{CPU} component.
}
\label{t05-2-1}
\begin{tabular}{lccccccc}
\toprule
\multirow{2}{*}{CaSE variants} & \multicolumn{7}{c}{MS MARCO test}                                                                                  \\
\cmidrule(l){2-8} 
                         & R-1            & R-2            & R-L            & BLEU           & MAP            & Recall@5        & NDCG        \\
                         \midrule
\ac{CaSE}                      & \textbf{57.44} & \textbf{39.26} & \textbf{49.91} & \textbf{32.08} & \textbf{65.75} & \textbf{92.20} & \textbf{74.34}
\\ 
CaSE-RG                   & 57.21          & 38.91          & 49.62          & 31.58          & 65.16          & 91.95          & 73.90   
\\
CaSE-STI                   & 55.68          & 36.94          & 47.92          & 29.98          & 64.02          & 91.10          & 73.03          
\\
CaSE-RPS                   &       56.36         &      38.37          &    48.75            &       31.22         & --              & --              & --
\\
\bottomrule
\end{tabular}
\end{table*}

\begin{table*}[t]
\caption{Ablation study on the SaaC test set (\%).
CaSE-X denotes CaSE with the component X left out.
CaSE-RG: replacing the \ac{RG} component with a traditional pointer generator.
CaSE-STI: \ac{CaSE} without the \ac{STI} component.
CaSE-RPS: \ac{CaSE} without the \ac{RPS} module. 
There is no CaSE-CPU because all other components rely on the \ac{CPU} component.
}
\label{t05-2-2}
\begin{tabular}{lccccccc}
\toprule
\multirow{2}{*}{CaSE variants}                                                          & \multicolumn{7}{c}{SaaC test}                                                                                        \\
\cmidrule(l){2-8}
                         & R-1            & R-2            & R-L            & BLEU           & MAP            & Recall@5        & NDCG             \\
                         \midrule
\ac{CaSE}                      & 37.34 & 14.58 & 27.24 & 13.26 & \textbf{17.22} & \textbf{13.72} & \textbf{32.67}
\\ 
CaSE-RG                   & \textbf{38.32}          & \textbf{15.20}          & \textbf{28.21}          & \textbf{14.45}          & 16.54          & 12.47          & 32.51          
\\
CaSE-STI                   & 37.74          & 14.51          & 27.62          & 13.01          & 16.70          & 12.68          & 32.39          
\\
CaSE-RPS                   &    37.83            &        15.02        &         26.87       &        14.39        & --              & --              & --              
\\
\bottomrule
\end{tabular}
\end{table*}

Generally, removing any module will result in a drop in performance in terms of both response generation and relevant passage selection.
Specifically, the results drop by more than 2\%pt in terms of BLEU and more than 1.5\%pt in terms of MAP on the MS MARCO dataset by removing the \ac{STI} module.
This is also true for the \ac{RPS} and \ac{RG} modules, although the drops are not as large as for \ac{STI}.
The generation performance of CaSE-STI is even worse than Masque, which confirms the effectiveness of \ac{STI}.
Interestingly, although we found that modeling passage selection is very important (Table~\ref{t05-1-1} and \ref{t05-1-2}) for the other models, removing the \ac{RPS} module does not much influence the overall performance of \ac{CaSE}.
We think the reason is that \ac{CaSE} incorporates the \ac{STI} module which has some overlapping effects with \ac{RPS} to some extent, as when a passage contains more supporting tokens, it is more relevant in general. 
To sum up, even if the \ac{RPS}, \ac{STI} and \ac{RG} share some common effects, they are also complementary to each other as combining them will bring further improvements.

One exception is that CaSE-RG achieves better response generation performance than \ac{CaSE} on \ac{SaaC}.
We believe that the reason for this behavior is that although \ac{CaSE} can generate more accurate responses by leveraging the outputs from the \ac{RPS} and \ac{STI} modules as priors in \ac{PPG} (which can be verified by the better performance on MS MARCO), this will influence the abstractiveness of the response, because \ac{CaSE} is encouraged to put more emphasis on the tokens in the relevant passages with \ac{PPG} and 
directly copy them to the 
responses instead of generating them from the vocabulary.
As a result, \ac{CaSE} performs better on the MS MARCO dataset while CaSE-RG performs better on the \ac{SaaC} dataset, because \ac{SaaC} is more abstractive, as can be seen from Table~\ref{t04-1}.


\subsection{Is there Room for further Improvement?}\label{s05-3}


To answer Q3, we explore the room for further performance improvements on the \ac{SaaC} dataset by conducting additional experiments and/or case studies.

\begin{table}[h]
\caption{Response generation and passage ranking performance of \ac{CaSE} on the \ac{SaaC} dataset w.r.t. different conversational turns (\%).}\label{t05-3}
\setlength{\tabcolsep}{2pt}
\begin{tabular}{cccccccc}
\toprule
Turn & R-1 & R-2 & R-L & BLEU  & MAP    & Recall@5 & NDCG \\
\midrule
1    &     \textbf{47.97}    &    \textbf{28.13}     &     \textbf{38.22}    &   \textbf{33.16}  & \textbf{0.2117} & \textbf{0.1909}   & \textbf{0.3521}   \\
2    &     38.05    &    16.27     &    27.00     &  15.56   & 0.1607 & 0.1043   & 0.3205   \\
3    &    42.11     &    17.25     &    30.54     & 13.95  & 0.1904 & 0.1289   & \textbf{0.3541}    \\
4    &     37.84    &     13.61    &    26.08     & 13.45  & 0.1635 & 0.1234   & 0.3127   \\
5    &    36.28     &    11.80     &     26.74    &  10.39   & 0.1406 & 0.0944   & 0.2992      \\
6    &    31.73     &    08.33     &     21.92    &  04.53   & 0.1361 & 0.0930   & 0.2950    \\
7    &     30.42    &    08.32     &     21.95    &   05.04   & 0.1451 & 0.1120   & 0.3000    \\
8    &     33.05    &     11.03     &    24.39     &   08.46   & 0.1060 & 0.0936   & 0.2415    \\
$\geq$ 9    &    35.16     &     13.88    &    26.06     &  11.00   & 0.1672 & 0.1088   & 0.3445     \\
\bottomrule
\end{tabular}
\end{table}
\begin{table}[h]
\caption{Response generation and passage ranking performance of Masque on the \ac{SaaC} dataset w.r.t. different conversational turns (\%).}\label{t05-3_}
\setlength{\tabcolsep}{2pt}
\begin{tabular}{cccccccc}
\hline
Turn     & R-1                             & R-2                             & R-L                             & BLEU                            & MAP                              & Recall@5                         & NDCG                             \\ \hline
1        & \textbf{47.42} & \textbf{27.02} & \textbf{36.55} & \textbf{28.49} & \textbf{0.2039} & \textbf{0.1550} & \textbf{0.3485} \\
2        & 30.99                           & 11.82                           & 21.70                           & 11.52                           & 0.1670                          & 0.1182                           & 0.3174 \\
3        & 34.94                           & 13.38                           & 24.40                           & 09.41                           & 0.2162                           & \textbf{0.1607}                           & \textbf{0.3689}                           \\
4        & 39.87                           & 17.51                           & 29.34                           & 17.41                           & 0.1536                           & 0.1114                           & 0.3039                           \\
5        & 35.85                           & 10.53                           & 27.20                           & 06.40                           & 0.1628                           & 0.0977                           & 0.3272                           \\
6        & 34.04                           & 11.51                           & 24.00                           & 11.42                           & 0.1479                           & 0.1423                           & 0.3115                           \\
7        & 31.26                           & 09.86                           & 22.55                           & 08.05                           & 0.1449                           & 0.1155                           & 0.3143                           \\
8        & 29.84                           & 06.95                           & 20.04                           & 05.12                           & 0.0754                           & 0.0339                           & 0.2158                           \\
$\geq$ 9 & 32.38                           & 10.02                           & 23.13                           & 07.27                           & 0.1624                           & 0.1279                           & 0.3330                           \\ \hline
\end{tabular}
\end{table}

First, more effort is needed to properly model the context, i.e., the conversational history, of the current turn query.
To illustrate this, we show the performance of response generation and passage ranking of each turn in Table~\ref{t05-3} and \ref{t05-3_}.
We found that \ac{CaSE} is better than Masque in general on the \ac{SaaC} dataset mostly because \ac{CaSE} performs better for the first three turns, and the improvements come mainly from the generation part.
This actually means that the proposed \ac{STI} and \ac{RPS} modules do indeed help to generate better responses, as without the two modules \ac{CaSE} basically degenerates into Masque.
However, it is worth to note that the advantage of \ac{CaSE} fades away as the conversational turns go on, which means that the main bottleneck becomes how to better model the context.

We see that the performance is much higher in the first turn because there is no context that needs to be considered.
Performance drops dramatically for the following turns including the 2-nd turn, and performance tends to get worse as the number of turns increases.
There is an exception for the $\geq$9-th turns, which are better than the 8-th turn in terms of passage ranking.
This may be because hard queries do not go beyond the 8-th turn in CAsT.
We analyzed the queries from the $\geq$9-th turns and found that there are a lot of ``what'' queries in these two turns like ``What type does chemical energy belong to?'', which generally needs less modeling of missing context.
A deeper understanding of the current query is challenging because it is not just a matter of coreference resolution~\citep{AliannejadiChiir20}.
It is common that people omit information to keep the conversations natural, which is well reflected in the \ac{SaaC} dataset.

Second, more effort is needed to obtain a better understanding of the current turn query.
To illustrate this, we conduct a comparison of using the original current query (OQ), the context queries and the current original query (CQ+OQ), and the reformulated current query (RQ), as shown in Table~\ref{t05-5}.
\begin{table}[t]
\caption{Response generation performance of \ac{CaSE} on the \ac{SaaC} dataset (\%).
OQ: Using the original query of the current turn; CQ+OQ: Using context queries + original query of the current turn; RQ: Using reformulated ground truth query of the current turn; GP: Using ground truth passages; GS: Using ground truth supporting sentences.
}
\label{t05-5}
\begin{tabular}{lcccc}
\toprule
            & R-1 & R-2 & R-L & BLEU  \\
            \midrule
OQ         &    39.24     &    18.71     &     29.49    & 18.05         \\
CQ+OQ &    37.34     &     14.58    &     27.24    &  13.26        \\
RQ     &    42.79     &     22.17    &     32.75    &   22.05      \\
\midrule
GP     &    46.19     &    25.78     &      33.37   &   27.05       \\
GS     &     48.74    &   29.39      &     36.54    &   32.34      \\
\bottomrule
\end{tabular}
\squeeze\squeeze
\end{table}
RQ is a manually rewritten version of the current original query, where we make sure that RQ is self-contained and does not need to rely on context.
We can see that using RQ improves the performance, which is not surprising.
But we should also note that even when using RQ, the performance is far from perfect as it does not even outperform the 1st turn in Table~\ref{t05-3}.
This indicates that the model cannot create a understanding of the current query even though all needed information is provided.
This is confirmed by the fact that CQ+OQ performs worse than OQ: the model is doing worse by involving more contextual information.

Third, more effort is needed to perform better at passage selection and support identification.
We illustrate this by showing the results of \ac{CaSE} when the ground truth passages (GP) or sentences (GS) are provided in Table~\ref{t05-5}.
We can see that although \ac{CaSE} has used effective and complex mechanisms to perform relevant passage selection and supporting token identification, there is still a lot of room to improve in this direction.

Fourth, more effort is needed to investigate how to generate more conversational and abstractive responses.
Due to limitations of the MS MARCO data, the models are rarely trained to learn to generate tokens that address the conversational nature of responses.
For instance, given the query ``What is arnica used for?,'' the current model will just list the answers, ``arnica , trauma , pain and shock.'' The human response is ``Arnica is a plant based remedy used to relief pain. It is also used to speed injury and trauma healing as well as to reduce bruising.''
Besides, there are some cases where there is no answer or only a partial answer is available in the passages; the current model will either generate a wrong answer or just leave it blank.
However, in practical scenarios, the system should indicate it does not know the answer or only knows part of the answer and reply accordingly, e.g., ``Sorry, I don't know much about the largest tiger shark ever to have lived on Earth or caught. But I do know the largest great white shark ever caught on camera, it was a seven meter-long female, called Deep Blue.''
We can build suitable datasets to address this.
Alternatively, we can investigate how to leverage datasets from related tasks, e.g.,  chitchat datasets, where there are more natural and conversational human responses~\citep{dinan2018wizard, moghe2018towards}.

\subsection{Case study}
\label{s05-4}
In order to obtain an understanding of what the responses look like, in this subsection we perform a qualitative analysis by examining some generated responses from CaSE and Masque.
We present 5 examples from the MS MARCO test set as shown in Table~\ref{t05-6-1}.
We also show 2 conversation examples from the SaaC test set as shown in Table~\ref{t05-6-2} and \ref{t05-6-3}.
We use the following ratings to indicate the quality of each response based on the ground truth responses.
++: very good (The generated responses are fluent, coherent, and contain all relevant information as in ground truth responses);
+: good (The generated responses are mostly fluent, coherent, and contain most relevant information as in ground truth responses);
0: fair (The generated responses are readable, and contain some relevant information as in ground truth responses);
-: bad (The generated responses are readable but contain wrong information);
{-}{-}: very bad (The generated responses are not readable at all or contain irrelevant information only).

\begin{table*}[]
\caption{Some examples from the MS MARCO test set.
}
\label{t05-6-1}
\begin{tabular}{@{}llm{10.5cm}m{0.5cm}@{}}
\toprule
\multirow{7}{*}{\rotatebox{90}{Example 1}} & Query  & cell organelles definition\\ 
                           & Human  & cell organelles is a membrane bound compartment or structure in a cell that performs a special function .                                                            &               \\
                           & CaSE   & a membrane - bound compartment or structure in a cell that performs a special function .                                                                             & +  \\
                           & Masque & cell organelles are a membrane - bound compartment or structure in a cell that performs a special function .                                                         & ++  \\ 
\midrule
\multirow{6}{*}[-0.1cm]{\rotatebox{90}{Example 2}} & Query  & what music style developed from scott joplin ? \\
                           & Human  & scott joplin developed the maple leaf rag and the entertainer .                                                                                                      &               \\
                           & CaSE   & the maple leaf rag and the entertainer .                                                                                                                             & +  \\
                           & Masque & the music of ragtime , scott joplin was the foremost composer of the genre in the early 20th century , known for works like the maple leaf rag and the entertainer . & +    \\
\midrule
\multirow{5}{*}[-0.2cm]{\rotatebox{90}{Example 3}} & Query  & glioma meaning                                                                                                                                                       &               \\
                           & Human  & the meaning of glioma is a tumor springing from the neuroglia or connective tissue of the brain , spinal cord , or other portions of the nervous system .            &               \\
                           & CaSE   & a glioma is a type of tumor that starts in the brain or spine .                                                                                                      & 0   \\
                           & Masque & a glioma is a type of tumor that starts in the brain or spine .                                                                                                      & 0   \\
\midrule
\multirow{7}{*}{\rotatebox{90}{Example 4}} & Query  & why does steam hammer occur ?                                                                                                                                        &               \\
                           & Human  & it occur by the accumulation of condensate ( water ) trapped in a portion of horizontal steam piping .                                                               &               \\
                           & CaSE   & because of the accumulation of condensate ( water ) trapped in a portion of horizontal steam piping .                                                                & ++    \\
                           & Masque & because it is caused by a steam bubble forming or being pushed into a pipe completely filled with water .                                                            & {-}{-}   \\
\midrule
\multirow{4}{*}[-0.2cm]{\rotatebox{90}{Example 5}} & Query  & how wide is a two way road ?                                                                                                                                         &               \\
                           & Human  & a two way road is between 44 and 48 feet wide .                                                                                                                     &               \\
                           & CaSE   & a two way road is between 44 and 48 feet wide .                                                                                                                      & ++    \\
                           & Masque & a two way road is 50 or 60 feet wide .                                                                                                                               & - \\ \bottomrule
\end{tabular}
\end{table*}

From Table~\ref{t05-6-1}, we can see that both CaSE and Masque perform well in general on the MS MARCO test set.
In most cases, they can find the most relevant passages, based on which they can generate proper responses.
The generated responses are fluent and coherent with the queries.
For example, when it is a `why' question, both models generate responses starting with `because'.
We can also see that CaSE is indeed better than Masque generally.
This is mostly because that CaSE is better at selecting the more relevant passages.
For instance, in Example 4 and 5, CaSE generates perfect responses by identifying the most relevant passages and supporting tokens while Masque generates improper or wrong responses because of the selection of the irrelevant passages.
This indicates that it is effective that CaSE uses the proposed \ac{STI} module and its corresponding loss to help better select the relevant passage and further identify the supporting tokens.
However, there are also cases when both models select the secondary relevant or irrelevant passages instead of the most relevant ones.
For instance, in Example 3, both models generate the same response which tries to explain the meaning of glioma.
Although the generated response is acceptable, it is not as good as the human-written response where more aspects about glioma are covered.

\begin{table*}[]
\caption{An example for the topic ``forms of energy'' from the SaaC test set.
}
\label{t05-6-2}
\begin{tabular}{@{}llm{10.5cm}m{0.5cm}@{}}
\toprule
\multirow{4}{*}[-1.5cm]{Turn 1} & Query  & what are the different forms of energy ?                                                                                                                                                                                                      &               \\
                        & Human  & as far as i know , the different forms of energy are kinetic energy , potential energy , gravitational energy , nuclear energy , chemical energy , heat energy , electrical energy , electromagnetic energy , sound energy and solar energy . &               \\
                        & CaSE   & light energy , heat energy , mechanical energy , gravitational energy , electrical energy , sound energy , chemical energy , nuclear or atomic energy .                                                                                       & +  \\
                        & Masque & light energy , heat energy , mechanical energy , gravitational energy , electrical energy , sound energy , chemical energy , nuclear or atomic energy .                                                                                       & +  \\
\midrule
\multirow{4}{*}[-0.5cm]{Turn 2} & Query  & how can it be stored ?                                                                                                                                                                                                                        &               \\
                        & Human  & energy can be stored in many ways like a gallon of gasoline or a barrel of oil contains stored energy that can be released when we burn it .                                                                                                  &               \\
                        & CaSE   & the different forms of energy can be stored in the guelph ultra store .                                                                                                                                                                       & - \\
                        & Masque & the different forms of energy can be stored in the form of glycogen in the liver and muscles , to be available for later use .                                                                                                                & - \\
\midrule
\multirow{4}{*}{Turn 3} & Query  & what type of energy is used in motion ?                                                                                                                                                                                                       &               \\
                        & Human  & the energy of motion is kinetic energy .                                                                                                                                                                                                      &               \\
                        & CaSE   & kinetic energy is the energy of motion .                                                                                                                                                                                                      & ++    \\
                        & Masque & the different forms of energy is stored in the kinetic energy .                                                                                                                                                                               & {-}{-}   \\
\midrule
\multirow{4}{*}[-0.5cm]{Turn 4} & Query  & tell me about mechanical energy .                                                                                                                                                                                                             &               \\
                        & Human  & mechanical energy can be either kinetic energy ( energy of motion ) or potential energy ( stored energy of position ) .                                                                                                                       &               \\
                        & CaSE   & mechanical energy can be either kinetic energy ( energy of motion ) or potential energy ( stored energy of position ) .                                                                                                                       & ++    \\
                        & Masque & mechanical energy is the energy which is possessed by an object due to its motion or due to its position .                                                                                                                                    & 0   \\
\midrule
\multirow{4}{*}[-1cm]{Turn 5} & Query  & give me some examples .                                                                                                                                                                                                                       &               \\
                        & Human  & examples of mechanical energy are the power of a football flying through the air or the wrecking ball that is swung backward away from a building .                                                                                           &               \\
                        & CaSE   & an example of mechanical energy is the power of a football flying through the air .                                                                                                                                                           & +  \\
                        & Masque & mechanical energy is the sum of kinetic and potential energy in an object that is used to do work .                                                                                                                                           & {-}{-}  \\
\bottomrule
\end{tabular}
\end{table*}

Table~\ref{t05-6-2} shows a conversation example on the topic ``forms of energy'' where CaSE performs well.
In most turns, CaSE can generate proper responses which means that CaSE can understand the queries and select the relevant passages.
Especially, compared with Masque, CaSE is better at handling queries at later turns.
For instance, in Turn 3, it seems that Masque is confused with the queries in Turn 2 and 3 and the generated response is a mixture of the answer to both Turn 2 and 3.
As on the MS MARCO test set, CaSE performs better in terms of selecting the relevant passages and identifying the supporting tokens.
This is illustrated in Turn 4, where CaSE selects the most relevant passage and generates the same response as the human response by identifying the right supporting tokens.
In contrast, the response generated by Masque is acceptable but not as good as CaSE.
The main reason is still that Masque selects the secondary relevant passage. 

\begin{table}[t]
\caption{An example for the topic ``Bronze Age collapse'' from the SaaC test set.
}
\label{t05-6-3}
\begin{tabular}{@{}llm{10.5cm}m{0.5cm}@{}}
\toprule
\multirow{4}{*}[-1cm]{Turn 1} & Query  & tell me about the bronze age collapse .                                                                                                                                                                                            &             \\
                        & Human  & it may be because of a shortage of tin , that is necessary for the melt of bronze that forced to seek an alternative this metal .                                                                                                  &             \\
                        & CaSE   & robert drews describes the collapse as " the worst disaster in ancient history , even more calamitous than the collapse of the western roman empire . "                                                                            & {-}{-} \\
                        & Masque & robert drews describes the collapse as " as 230 years , corresponding to the uncertainty regarding the length of the " dark age " of mesopotamia with regard to the uncertainty regarding the length of the " .                    & - \\
\midrule
\multirow{4}{*}[-0.5cm]{Turn 2} & Query  & what is the evidence for it ?                                                                                                                                                                                                      &             \\
                        & Human  & could be the extrinsic evidence , but i am not sure whether it can support the collapse of the bronze age .                                                                                                                        &             \\
                        & CaSE   & the bronze age collapse .                                                                                                                                                                                                          & {-}{-} \\
                        & Masque & the bronze age collapse and trade disruptions in the mediterranean around 1300 bc , forced metalworkers to seek an alternative to bronze .                                                                                         & {-}{-} \\
\midrule
\multirow{4}{*}[-1cm]{Turn 3} & Query  & what are some of the possible causes ?                                                                                                                                                                                             &             \\
                        & Human  & one of the possible causes of the bronze age collapse is the invasion of the sea peoples .                                                                                                                                         &             \\
                        & CaSE   & the iron age in europe is being seen as a part of the bronze age collapse in the ancient near east , in ancient india ( with the post - rigvedic vedic civilization ) , ancient iran ,                                             & {-}{-} \\
                        & Masque & the greek tradition relates this destruction to the dorians , but it is suggested that the dorian invasion was only one of the causes of the bronze age collapse in the eastern mediterranean , as there is no evidence            & + \\
\midrule
\multirow{4}{*}[-1.5cm]{Turn 4} & Query  & who were the sea peoples ?                                                                                                                                                                                                         &             \\
                        & Human  & the sea peoples was a confederacy of seafaring raiders that caused political unrest , and attempted to enter or control egyptian territory during the late 19th dynasty , and the 20th dynasty .                                   &             \\
                        & CaSE   & the bronze age collapse is the transition from the late bronze age to the early iron age , expressed by the collapse of palace economies of the aegean and anatolia , which were replaced after a hiatus by the isolated           & {-}{-}  \\
                        & Masque & the sea peoples was a confederacy of seafaring raiders who sailed into the eastern shores of the mediterranean , caused political unrest , and attempted to enter or control egyptian territory during the late 19th dynasty , and & + \\
\midrule
\multirow{4}{*}[-0.5cm]{Turn 5} & Query  & what was their role in it ?                                                                                                                                                                                                        &             \\
                        & Human  & the sea peoples entered and invaded eastern mediterranean territory . the sea people ' s invasions ushered or caused the bronze age collapse .                                                                                     &             \\
                        & CaSE   & the sea peoples ushered the bronze age collapse , which resulted in the cultural collapse of mycenean greece , the hittite empire , the new kingdom of egypt and the civilizations of canaan and syria .                                                                                                                                                                             & + \\
                        & Masque &  robert drews describes the collapse as " " " " " " " .                          & {-}{-}\\
\bottomrule
\end{tabular}
\end{table}

It should be noted that there are still some cases where both CaSE and Masque perform badly.
An example on the topic ``Bronze Age collapse'' is shown in Table~\ref{t05-6-3}.
We can see that, for most turns, both CaSE and Masque generate unsatisfactory responses.
The reasons for the poor performance are three-fold.
First, the topics are more difficult in that the passage is longer and there are more confusing concepts.
For instance, for Turn 1, there are 71 passages annotated as relevant but only 3 of them are annotated as most relevant.
A passage is annotated as most relevant only if it contains all the relevant information to form a complete answer.
The most relevant passages contains the complete answers
As a result, it is difficult for the models to find the passage that contains the answers from a couple of passages which are similar but do not contain the answers or only contain part of the answers.
This can be further confirmed by some of the statistics, e.g., `\#passage length', `\#pairwise passage similarity' in Table~\ref{t04-1}.
Second, the follow-up queries depend more on the previous queries and modeling of long-term dependency might be needed.
For comparison, for the example in Table~\ref{t05-6-3}, in order to understand the queries in later turns (Turn 4 and 5), the approaches still need an understanding of the background `Bronze Age collapse' from the the query.
Third, there are cases where no candidate passage contains a direct answer but just some secondary relevant information instead.
For example, we found that, in the SaaC dataset, there are many human-written responses containing statements like ``could be'', ``I think'', ``I am not sure'', or even ``I don't know'', which indicate that when the secondary relevant passages are provided, even for humans, it is sometimes not easy to write the responses.
Finally, the above analysis also indicates that the \ac{SaaC} dataset brings new challenges that remain to be solved in the future. 

%% file: sections/06-relatedwork.tex

\section{Related Work}
\label{section:RelatedWork}

We briefly present an overview of related work on \acf{CS} and on \acfp{CA}.

\subsection{Conversational Search}

The concept of search as a conversation has been around since the 1980s~\cite{belkin1980anomalous,Croft:1987:IRN:35053.35054}.
%
Until recently, the idea did not attract a lot of attention due to limitations in data and computing resources at the time.
Now, the topic is back in the spotlight.
One branch of work conducts user studies on \ac{CS}.
For example, \citet{Vtyurina:2017:ECS:3027063.3053175} conduct a user study, where they ask 21 participants to solve 3 information seeking tasks by conversing with three agents: an existing commercial system, a human expert, and a perceived experimental automatic system, backed by a human ``wizard behind the curtain.''
They show that existing conversational assistants cannot be effectively used for complex information search tasks.
\citet{vakulenko-conversational-2017} argue that existing studies neglect exploratory search when users are unfamiliar with the domain of their goal.
They investigate interactive storytelling as a tool to enable exploratory search within the framework of a conversational interface. 
\citet{Trippas:2018:IDS:3176349.3176387} conduct a laboratory-based observational study for \ac{CS}, where pairs of people perform search tasks communicating verbally.
They conclude that \ac{CS} is more complex and interactive than traditional search.

Another line of work has proposed theoretical frameworks concerning \ac{CS}.
\citet{Radlinski:2017:TFC:3020165.3020183} present a theoretical framework of information interaction in a chat setting for \ac{CS}, which highlights the need for multi-turn interactions.
\citet{fa879899b8a74e57a989884eefa4730e} propose a conceptual framework that outlines the actions and intents of users and agents in order to enable the user to explore the search space and resolve their information need.
The work listed above studies \ac{CS} either in a theoretical or a user study environment.
The theoretical/conceptual frameworks have made requirements about the data annotations more demanding, often going beyond currently available datasets.

\citet{Zhang:2018:TCS:3269206.3271776} devise a System Ask-User Respond paradigm for \ac{CS}, and design a memory network for product search and recommendation in e-commerce.
\citet{lei2020estimation} initiate a path breaking Estimation—Action—Reflection framework to achieve deep interaction between recommendation and conversation~\cite{lei2020interactive}. 
Following this line, \citet{Aliannejadi:2019:ACQ:3331184.3331265} and \citet{Hamedwww2020} formulate the task of asking clarifying questions in information retrieval.
%
\citet{Bi:2019:CPS:3357384.3357939} propose a conversational paradigm for product search, and an aspect-value likelihood model to incorporate both positive and negative feedback on non-relevant items.
\citet{macaw} introduce an open-source framework with a modular architecture for \ac{CS}. 
It has a modular design to encourage the study of new \ac{CS} algorithms.
It can also integrate with a user interface to facilitate user studies.
To advance research on \ac{CS}, a number of \ac{CS} datasets have been released recently.\footnote{https://github.com/chauff/conversationalIR}
For example, \citet{penha2019mantis} release a multi-domain dialogue dataset containing information-seeking conversations from the Stack Exchange community.\footnote{https://stackexchange.com/}
\citet{cast2019} organize a TREC Conversational Assistance Track (CAsT), which establishes a concrete and standard collection of data with information needs to make systems directly comparable.
In the first year, the benchmark only focuses on candidate passage retrieval: Read the dialogue context and perform retrieval over a large collection of passages.

Although the studies listed above propose concrete datasets or methods for \ac{CS}, none of them targets directly generating conversational system responses by modeling \ac{CS} in a search engine scenario.
In contrast, instead of just retrieving relevant passages, we propose to directly generate conversational responses for \ac{CS}, which is more challenging as we need to not only find the relevant passage but also summarize it into a short conversational response.
Note that although summarization is involved in this process, it is quite different from query-driven summarization~\cite{10.1145/3077136.3080792}, where: 
\begin{enumerate*}
\item the queries are more like a general information need description, e.g., the query ``Describe steps taken and worldwide reaction prior to introduction of the Euro on January 1, 1999. Include predictions and expectations reported in the press.'' from DUC 2007;\footnote{\url{https://duc.nist.gov/duc2007/tasks.html}}
\item the passages are mostly from articles like news reports which are in formal language; and
\item the answer is usually much longer, e.g., on DUC 2007, answers can be as long as 250 words.
\end{enumerate*}
Recently, \citet{10.1145/3176349.3176388} and \citet{TRIPPAS2020102162} create the MISC dataset and the CSCData dataset for spoken conversational search by recording pairs of seeker and intermediary collaborating on information seeking.
However, there are no supporting documents available and the responses are in very verbose spoken language.
In most cases, the responses are more like a description of SERPs instead of a direct summary of the answers.

\subsection{Conversational Agents}

Conversational modeling has long been a hot research topic in natural language processing~\cite{lester2004conversational,higashinaka2014towards}.
Most research falls into three groups: \acf{TDS} agents~\cite{DBLP:conf/icassp/ShalyminovSAS20,DBLP:journals/corr/abs-2003-01680}, social bots, and \ac{QA} agents~\cite{gao-etal-2018-neural}.
\ac{TDS} aims to achieve a specific task for users, e.g., booking a flight~\cite{budzianowski-etal-2018-multiwoz,lei2018sequicity}, while social bots aim to satisfy the human need for communication and so on~\citep{dinan2018wizard,meng-2020-refnet}.
These goals are quite different from people's goals in search scenarios where user information needs can be more complex and exploratory.

Efforts to build \ac{QA} agents come in two main flavors: \ac{KBQA} and \ac{MRC}, which study how to query a KB interactively with natural language, and generate an answer to an input query based on a set of passages, respectively.
Existing methods are either retrieval/extraction-based~\cite{hermann2015teaching,das2018multistep} or generation-based~\cite{yin-etal-2016-neural-generative}.
The former try to query an entity from a KB~\cite{yih2016value}, retrieve evidence from evidence candidates~\cite{lee-etal-2019-latent}, or extract a span from a given passage as direct answers~\cite{rajpurkar-etal-2016-squad}.
The latter generates natural language answers directly~\cite{Tan2018snet}.
Recently, \ac{QA} has been extended to multi-turn conversational scenarios~\cite{choi-etal-2018-quac}, which introduce more challenges related to conversational understanding.
For example, \citet{10.1145/3357384.3358016} target complex questions in KB-QA, which involve joining multiple relations and multi-hop reasoning in a conversational setting~\cite{NIPS2018_7558, shen-etal-2019-multi}.
\citet{10.1145/3397271.3401110} study how to retrieve evidence from a large collection in a multi-turn conversational context.
On many benchmark datasets, the best models are approaching human performance~\citep{DBLP:conf/aaai/SahaPKSC18,choi-etal-2018-quac} or even surpass humans~\citep{reddy-etal-2019-coqa}.
However, the studies listed above mostly target fact-centric information needs that can be answered by entity mentions or triples from a \ac{KB}~\cite{joshi-etal-2017-triviaqa}.
Besides, the passages are fixed for all conversation turns and from a single source, i.e., Wikipedia, which is different from search scenarios where \acp{SERP} are generated with information from multiple sources.
Some approaches target complex questions that cannot be well answered by \ac{KB} triples ~\cite{kocisky-etal-2018-narrativeqa,Qu:2018:ACU:3209978.3210124}.
In particular, \citet{DBLP:conf/nips/NguyenRSGTMD16} collect a large-scale dataset, MS MARCO, from Bing usage logs, where the answers are written by real humans to ensure that they are in natural language.

The work listed above focuses on single-turn \ac{QA}.
In this work, we extend it to multi-turn conversational scenarios where query understanding, relevant passage finding and response generation, etc. are more challenging.
In order to achieve better relevant passage finding and response generation, we further propose the \ac{CaSE} model, which improves a state-of-the-art model by introducing the \ac{STI} module and a weakly-supervised \ac{CCCE} loss.


%% file: sections/07-conclusion.tex

\section{Conclusion and Future Work}
\label{section:conclusion}

In this paper, we propose \emph{conversations with search engines} as task for the community to consider and we contribute two types of result:
First, we release a new test set, \ac{SaaC}, which is more suitable and challenging for this research than existing resources.
Second, we propose an end-to-end neural model, \ac{CaSE}, to advance the state-of-the-art.
We implement state-of-the-art methods from related tasks and conduct extensive experiments to show that: 
\begin{enumerate*}
\item the proposeed \ac{CaSE} model can achieve state-of-the-art performance;
\item the proposed \ac{STI} and \ac{PPG} modules can bring large improvements; and
\item \ac{SaaC} is a more challenging dataset than previously introduced ones, leaving significant room for further improvements.
\end{enumerate*}

A limitation of this work is that the \ac{SaaC} dataset is built upon the TREC CAsT dataset which adopts a ``user asks - system responds'' paradigm, and, hence, does not consider any mixed initiative aspects.
It is also worth mentioning that we assume that each conversation session is well separated in this work.
However, in a practical system, a component is needed to detect when the users start a new session.

As to future work, on the one hand, we plan to further improve the performance of \ac{CaSE} by proposing transfer learning methods in order to leverage more multi-turn conversational datasets (e.g., conversational \ac{QA}~\cite{yang2018hotpotqa,reddy-etal-2019-coqa}, conversational \ac{MRC}~\cite{choi-etal-2018-quac} and chitchat~\cite{lowe-etal-2015-ubuntu}).
On the other hand, we plan to study how to address the cases in \ac{SaaC} when there is no correct answer or only a partially correct answer.
Besides, it remains to be studied what progress we can make when different aspects are considered simultaneously in one system or model, e.g., asking clarifying questions~\cite{Aliannejadi:2019:ACQ:3331184.3331265,Hamedwww2020}, conversational search and recommendation~\cite{cast2019,10.1145/3397271.3401130,10.1145/2939672.2939746,10.1145/3209978.3210002}, chitchats~\cite{wu-yan-2018-deep}, conversation management~\cite{TRIPPAS2020102162} and so on.